\documentclass[pre,superscriptaddress,longbibliography]{revtex4-1}
\usepackage{graphicx,amssymb,amsmath}
\usepackage{xcolor}
\usepackage[applemac]{inputenc}
\usepackage{epstopdf}
\usepackage{float}

\begin{document}

\title{3-wave and 4-wave interactions in gravity wave turbulence}
\author{Quentin Aubourg}
\author{Antoine Campagne}
\affiliation{Laboratoire des Ecoulements G\'eophysiques et Industriels (LEGI), Universit\'e Grenoble Alpes, CNRS, F-38000 Grenoble, France}
\author{Charles Peureux}
\author{Fabrice Ardhuin}
\affiliation{Laboratoire d’Oc\'eanographie Physique et Spatiale (LOPS), Univ. Brest, CNRS, Ifremer, IRD, F-29200 Plouzan\'e, France}
\author{Joel Sommeria}
\author{Samuel Viboud}
\author{Nicolas Mordant}
\email[]{nicolas.mordant@univ-grenoble-alpes.fr}
\affiliation{Laboratoire des Ecoulements G\'eophysiques et Industriels (LEGI), Universit\'e Grenoble Alpes, CNRS, F-38000 Grenoble, France}

\begin{abstract}
The Weak Turbulence Theory is a statistical framework to describe a large ensemble of nonlinearly interacting waves. The archetypal example of such system is the ocean surface that is made of interacting surface gravity waves. Here we describe a laboratory experiment dedicated to probe the statistical properties of turbulent gravity waves. We setup an isotropic state of interacting gravity waves in the Coriolis facility (13~m diameter circular wave tank) by exciting waves at 1~Hz by wedge wavemakers. We implement a stereoscopic technique to obtain a measurement of the surface elevation that is resolved both in space and time. Fourier analysis shows that the laboratory spectra are systematically steeper than the theoretical predictions and than the field observations in the Black Sea by Leckler {\it et al. JPO} 2015. We identify a strong impact of surface dissipation on the scaling of the Fourier spectrum at the scales that are accessible in the experiments. We use bicoherence and tricoherence statistical tools in frequency and/or wavevector space to identify the active nonlinear coupling. These analyses are also performed on the field data by Leckler {\it et al.} for comparison with the laboratory data. 3-wave coupling are characterized and shown to involve mostly quasi resonances of waves with second or higher order harmonics. 4-wave coupling are not observed in the laboratory but are evidenced in the field data. We finally discuss temporal scale separation to explain our observations.
\end{abstract}

\maketitle

\section{Introduction}

The Weak Turbulence Theory (WTT) is a statistical theory that describes the evolution of an ensemble of interacting non-linear waves (wave turbulence). Applicable to a wide variety of dispersive waves (for reviews, see ~\cite{R1,R2,R3}), the theory has been historically developed in the 60's by Zakharov \cite{zakharov1984kolmogorov} in plasma physics and Hasselmann in oceanography for the modeling of surface gravity waves~\cite{R9}. The application of the theory relies on two asymptotic hypotheses: the domain must be large and non-linearities weak. Thus, a scale separation appears between the linear time of the wave $T^{L}=2\pi/\omega$ and the typical slow non-linear time $T^{NL}$ related to the non-linear waves interactions. As the non-linearity is weak, only resonant interactions are able to develop a long time coupling and then transfer energy. The resonant conditions impose that the wave vector $\bf k_i$ and the frequencies $\omega_i$ follow:
\begin{equation}
\mathbf{k_1}=\mathbf{k_2}+\mathbf{k_3} \quad\quad \omega_1=\omega_2+\omega_3
\label{eq3w}
\end{equation}
at the lowest order that involves $3$-waves interactions. The number $N$ of waves in an interaction is generally set by the order $N-1$ of the non-linear term in the dynamic equations. Thus, for surface water waves for which the non-linear term is quadratic, we expect $N=3$. However $N$ is also ruled by the existence or the absence of geometrical solutions to the resonant equations. For freely propagating waves, these solutions can be found using the linear dispersion relation of the waves. For surface gravity waves, the negative curvature of the dispersion relation $\omega=\sqrt{gk}$ does not allow for resonant waves with $N=3$, and $N=4$ should be considered.
\begin{equation}
\mathbf{k_1}+\mathbf{k_2}=\mathbf{k_3}+\mathbf{k_4} \quad\quad \omega_1+\omega_2=\omega_3+\omega_4\, .
\label{eq4w}
\end{equation}
In the following, when using the words ``N-wave resonance'', we mean a set of Fourier modes $(\omega,\mathbf k)$ that fulfills the resonance equations (\ref{eq3w}) or (\ref{eq4w}) independently of whether the Fourier mode corresponds to an actual freely propagating wave (ie following the linear dispersion relation) or not. In particular it may include bound waves generated by quadratic nonlinearities. This is somewhat different to the meaning used most often in the oceanographic community, namely what we call 3-wave interaction that involves a bound component is usually referred to as a 4-wave interaction.

The WTT relies on these resonant equations and allows to solve analytically the temporal evolution of statistical quantities in several physical systems. In the situation of an out of equilibrium and stationary forced system, it leads to the so-called Kolmogorov-Zakharov (KZ) power spectrum~\cite{R1}. For gravity waves, since the dispersion relation follows a unique power law, solutions can be found as well using dimensional analysis~\cite{R2}. It leads to the following prediction for the spectrum of the surface elevation $\eta$: $E^\eta(k)\propto g^{-1/2}P^{1/3}k^{-5/2}$ where $k=\left\|\mathbf{k}\right\|$, $g$ is the gravity and $P$ is the average injected power. The exponent $1/3=1/(N-1)$ is related to the $4$-wave resonance. Using the linear dispersion of surface gravity waves $\omega=\sqrt{gk}$, it is then possible to express the spectrum in the frequency space : $E^\eta(\omega)\propto gP^{1/3}\omega^{-4}$.

Although gravity waves are among the pioneering systems for the study of non-linear waves interactions, laboratory observations show a strong variability. Unlike capillary waves-assemblies for which the WTT predictions of the spectral exponent seem consistent with the observations~\cite{Berhanu2013,Falcon2007,Falcon2009,Deike2014,CoPr11}, spectra of pure gravity waves are often in contradiction with the WTT and show a forcing-dependent power spectrum. The spectral exponent predicted by the WTT is actually observed only for strongly forced systems, which is obviously inconsistent with the major hypothesis of weak nonlinearity~\cite{Nazarenko2009,Deike2015,Denissenko2007,Falcon2007}, whereas the exponent is significantly steeper than the prediction in the weak non-linearity limit. The main reason that is generally proposed to explain these discrepancies is a failure of the condition of an asymptotic large system or the presence of overturning waves. As experiments have always a finite size, a discretization of the $\mathbf{k}$-space appears and may not allow to satisfy the resonant conditions \cite{Kartashova1998}. These discrete effects can be overcome when non-linearities increase: non-linearity induces a finite spectral width of the modes and may allow quasi-resonant interactions between waves that are not strictly on the linear dispersion relation~\cite{Pushkarev1999,Nazarenko2013}. However, this situation is often in contradiction with the second hypothesis of weak non-linearities if too strong linearities are required to overcome the discreteness of the modes. A second cause for the observed differences with the WTT predictions may be attributed to the presence of dissipation in the inertial range where the energy cascade is operating. It generates a leak of energy and tends to steepen the power spectrum of the waves and thus decrease the spectral exponent as it has been observed for flexural wave turbulence~\cite{R23,Humbert}. It is well known that surface waves are very sensitive to interface pollution that forms a very thin layer of organic material at the surface (oil, surfactants,...). Due to the presence of an elastic pollution film at the surface, the stress boundary conditions are changed compared to the ideally clean case and the induced boundary layer is considerably more dissipative \cite{Alpers1989}. The pollution layer is deformed by fluctuations of the surface, and it is subject to compressive waves~\cite{Behroozi2007}.  These waves, referred to as Marangoni waves, are the main cause of dissipation of surface waves through the boundary layer with a maximal efficiency at frequencies close to $4$~Hz, at which a resonance appears between the dispersion relations of surface gravity waves and Marangoni waves~\cite{Alpers1989,Przadka2011}. The addition of these two effects are most likely responsible for the difference between the observed wave power spectrum and the WTT predictions. 

The goal of the present article is thus to investigate directly the wave interactions that are at the core of the theory. We perform an analysis of high order correlation similar to that presented in previous investigations of surface gravity-capillary wave turbulence ~\cite{Aubourg2015,Aubourg2016}. The experimental setup is presented in the part~\ref{ExpSet}. A first analysis of the power spectrum and the effect of the pollution is detailed in part~\ref{SA}. Then 3-wave and 4-wave resonant interactions are investigated in part~\ref{3w} and \ref{4w}. 

\section{Experimental setup}
\label{ExpSet}

\begin{figure}[!htb]
\includegraphics[width=14cm]{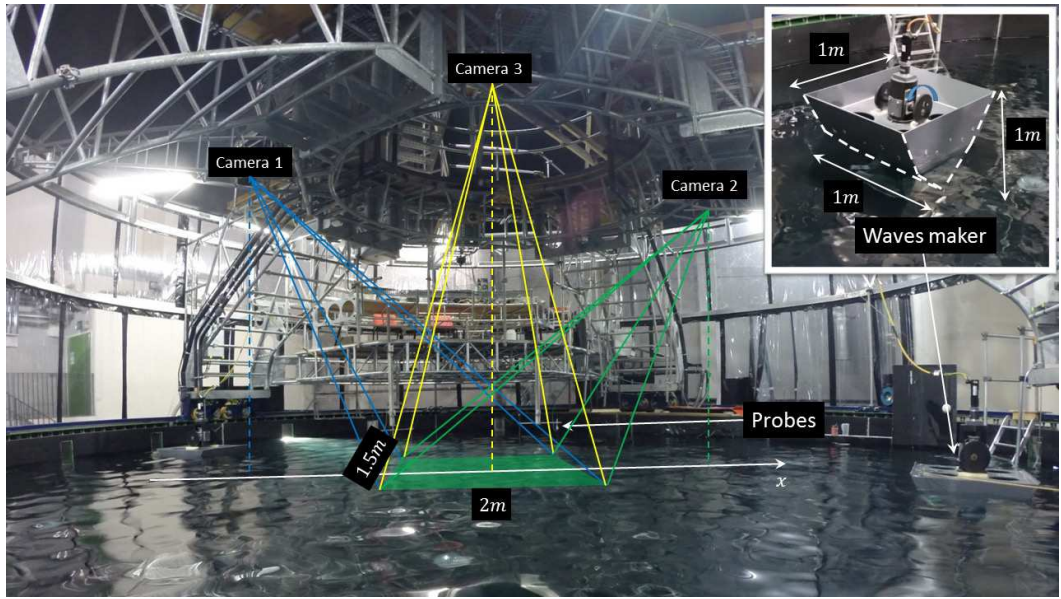}
\caption{Picture of the experimental setup : waves are generated by vertical oscillations of two wedge wavemakers of size $1\textrm{m}\times1\textrm{m}\times1\textrm{m}$ (visible in the inset). The deformation of the interface is measured in space and time using three cameras positioned above the surface. The common field of view of the cameras covers a surface of $2\times1.5$~m$^2$. We seed the surface with polystyrene buoyant particles of $0.7$~mm diameter. We perform cross-correlation between the different views in order to reconstruct the interface deformation $\eta(x,y,t)$ as well as the velocity field $\mathbf u=(u,v,w)$~\cite{AubourgPhD}. A set of four capacitive probes are also used to obtain single point measurements.}
\label{fig1}
\end{figure}

\begin{figure}[!htb]
\includegraphics[width=10cm]{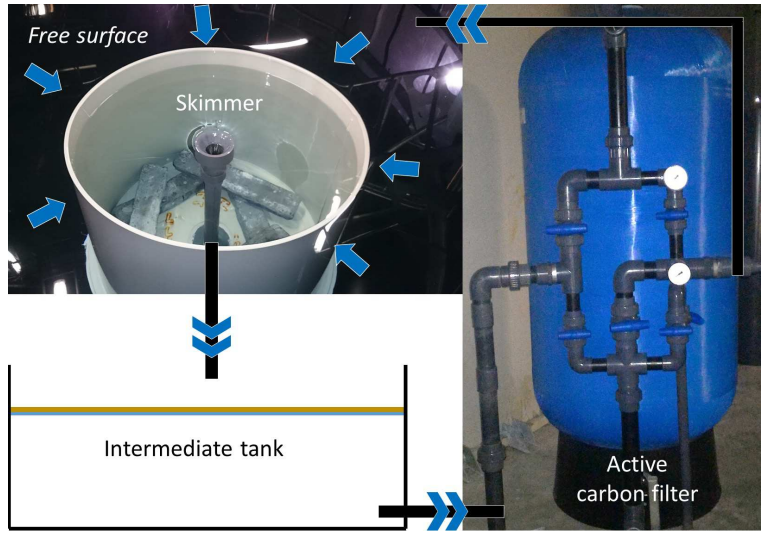}
\caption{Setup used to clean up the free surface: the upper water layer flows in a $1$~m cylinder located near the wall of the circular tank and connected to an intermediate tank located below the main one. The water is then pumped through an active carbon filter before being reinjected at the surface at a location diametrically opposite to the skimmer.}
\label{fig1b}
\end{figure}

Figure \ref{fig1} shows the general setup of the experiment. Gravity waves are generated in a $13$~m diameter circular tank filled with fresh water at a rest height equal to $70$~cm. In order to restrict surface pollution, the water surface is cleaned using a skimmer (see figure \ref{fig1b}): the upper water layer flows in a $1$~m cylinder connected to an intermediate tank located below the main one. The water is then pumped through a active carbon filter before being re-injected at the surface of the wavetank. The filtration is operated during a few hours in order to obtain the cleanest water surface. Waves are generated using two vertically oscillating wedge wavemakers of size of about $1\times1\times1 $~m$^3$ and placed at roughly $1.5$~m from the wall (Fig.~\ref{fig1}). They are driven within a range of $(2-4)$~cm in amplitude and with a randomly modulated oscillation frequency centered around $1$~Hz. The modulation follows a Gaussian distribution of width $0.15$~Hz. The curved walls provide an efficient mixing of the waves and a rather homogeneous state is observed. 

Two methods are implemented to measure the surface deformation: a local one with four capacitive probes and a time and space-resolved scheme using three cameras located about 4~m above the free surface. In the latter case, called Stereo-PIV in the following, buoyant particles are seeded at the surface in order to form a random pattern which is recorded simultaneously by the cameras. A cross-correlation between the three cameras at the same time is first computed to reconstruct the free surface using a stereoscopic algorithm. In a second step, a PIV (Particles Images Velocimetry) measurement is performed that provides the velocity field in the framework of each cameras. We finally merge all the PIV fields by using the stereoscopic reconstruction that allows us to combine correctly the velocity components of distinct cameras at the deformed surface of water. In this way we obtain a complete 2D resolved velocity field at the free surface (see \cite{AubourgPhD} for details of the image processing). In the ``classical'' Stereo-PIV technique the velocity is reconstructed in the bulk of the fluid in a flat plane illuminated by a laser sheet. Here the velocity field is computed at the surface of water which is not flat and it makes the reconstruction algorithm quite different. For simplicity we still refer to our method as Stereo-PIV. The main reason for implementing this global method instead of using only the stereo surface reconstruction is the improvement of the dynamical range of the measurement by measuring directly the power spectrum of the speed which is mathematically shallower than the one computed from the stereo (the former being the spectrum of a velocity and the latter of the elevation). We found a gain of about one order in magnitude in energy~\cite{AubourgPhD}. The knowledge of the velocity field also allows us to directly measure the non linear advective term as it will be presented below in part~\ref{SA}.

\begin{figure}[!htb]
\includegraphics[width=11cm]{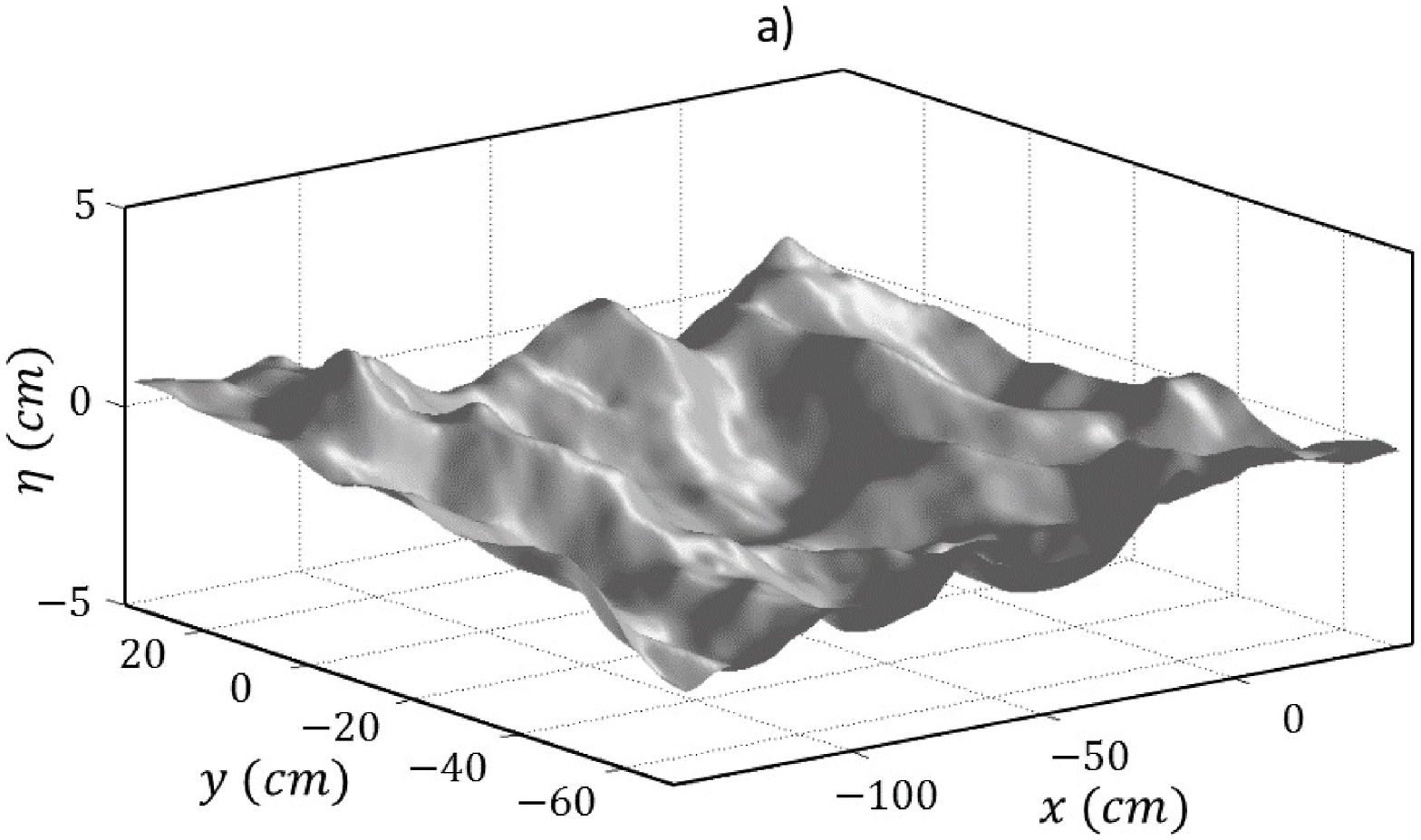}
\includegraphics[width=8cm]{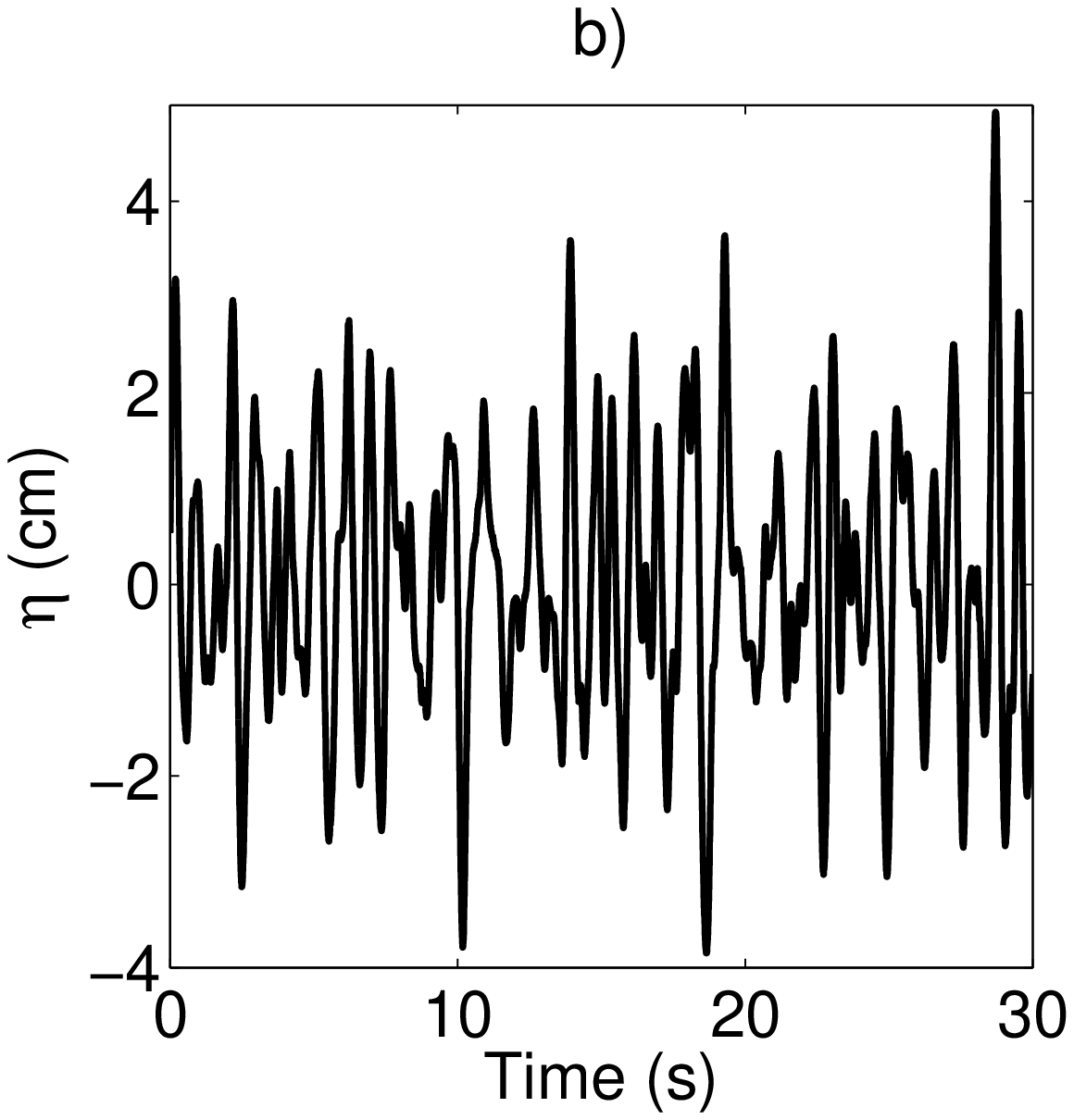}
\includegraphics[width=8cm]{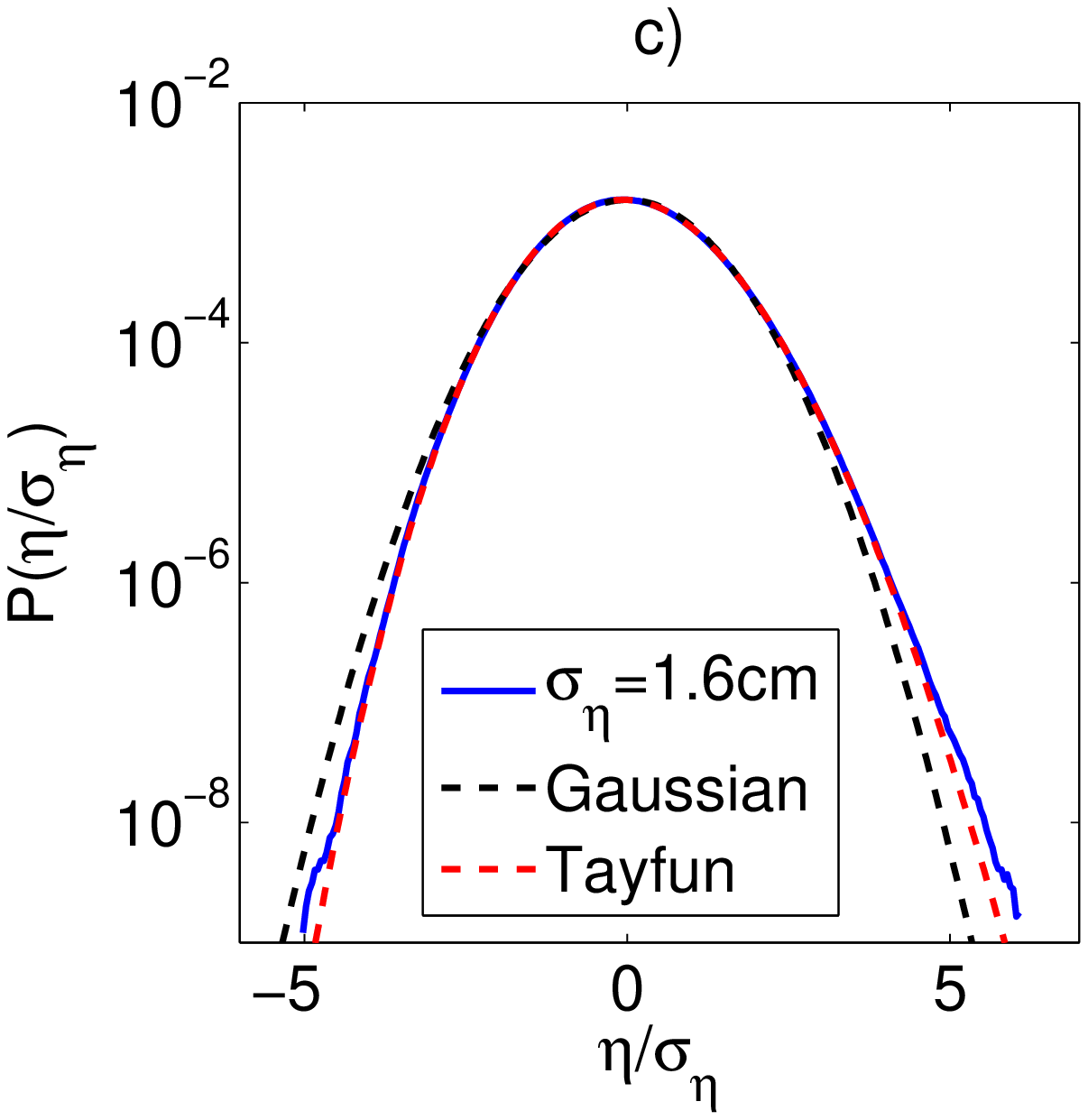}
\caption{(a) Snapshot of the 3D reconstruction of the free surface elevation $\eta(x,y,t)$. (b) Temporal evolution of $\eta$ at one given location in the image. (c) Probability density function of $\eta$ normalized by the standard deviation of the wave height $\sigma_\eta=1.6$~cm. The dashed black line is the normal distribution. The red dashed line is the Tayfun distribution which incorporates non-linear second order effects (see text).}
\label{fig2}
\end{figure}
The stereoscopic method allows us to reconstruct the free surface elevation $\eta$ over a field of view of about $2\times1$~m$^2$ with a vertical sensibility close to $0.6$~mm~\cite{AubourgPhD}. The spatial resolution is restricted by the inhomogeneous seeding of the particles. We used polystyrene particles with a diameter of $0.7$~mm. Due to well known capillary effects, a depression of the free surface is created between particles that generates an attraction that can act over a distance greater than $10$ diameters~\cite{Kralchevsky2000,Gifford1971}. For very weak waves, the particle dispersion by the waves is not efficient enough to balance this tendency to coalesce and thus we are not able to reduce the PIV correlation window below a size of $30$ pixel, which corresponds to a horizontal resolution of $4$~cm.

Figure~\ref{fig2}(a) shows a snapshot of the spatial reconstruction of $\eta$ and Fig.~\ref{fig2}(b) an example of the temporal evolution at one location. Figure~\ref{fig2}(c) displays the distribution of $\eta$ normalized by the standard deviation wave height $\sigma_{\eta}=\sqrt{\langle(\eta^2\rangle}=1.6$~cm (assuming $\langle \eta\rangle=0)$. We observe a slight deviation from the Gaussian distribution. Our observation is better described by the Tayfun distribution~\cite{Tayfun1980} simplified by Socquet-Juglard {\it et al.}~\cite{Socquet-Juglard2005} :
\begin{equation}
P(\eta)=\frac{1-\frac{7\sigma_{\eta}^2k_{p}^2}{8}}{\sqrt{2\pi(1+3G+2G^2)}}\exp(-\frac{G^2}{2\sigma_{\eta}^2k_{p}^2}) \quad\quad G=\sqrt{1+2k_{p}\sigma_{\eta}\eta}-1
\label{equtay}
\end{equation}
where $k_p$ is the wavenumber of the maximum peak in the power spectrum. This correction suggest the presence of harmonic waves (2nd order) over a Gaussian linear field as it is commonly observed in gravity wave experiments ~\cite{Denissenko2007,Onorato2009}. We thus suspect the presence of significant nonlinearities in our experiment.

\clearpage

\section{Spectral Analysis}

\label{SA}

\subsubsection{Single point spectra and water pollution}
\begin{figure}[!htb]
\includegraphics[width=14cm]{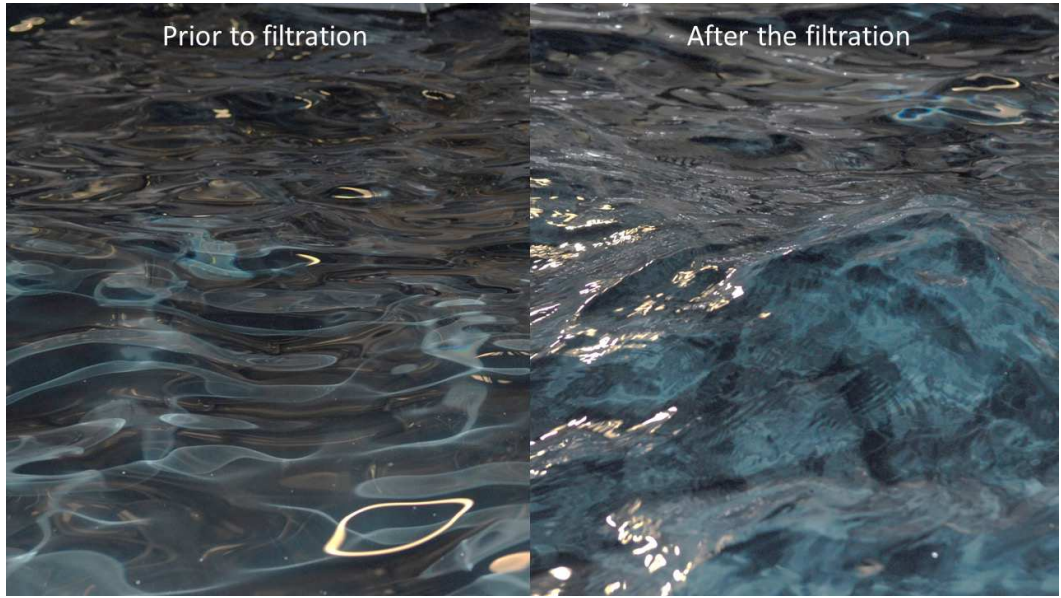}
\caption{Two pictures of the free surface taken prior to the filtration (left) and after filtering during a few hours. The intensity of the forcing is equivalent for both case. The filtration change significantly the surface and capillary waves become present.}
\label{fig3}
\end{figure}

We begin with the analysis of the data recorded by the capacitive probes in a large variety of regimes to investigate the influence of water pollution. A strong sensitivity to the filtering of the water surface has been found. Figure~\ref{fig3} compares two pictures of the free surface before and after the filtration under identical forcing conditions.
Before the filtration (Fig.~\ref{fig3}(a)), the free surface is smooth. This suggests the presence of a strong dissipation of short waves. Shortly after stopping the filtration (Fig.~\ref{fig3}(b)), the wave field looks very different with numerous short capillary waves riding on the crest of longer waves. 

\begin{figure}[!htb]
\includegraphics[width=8cm]{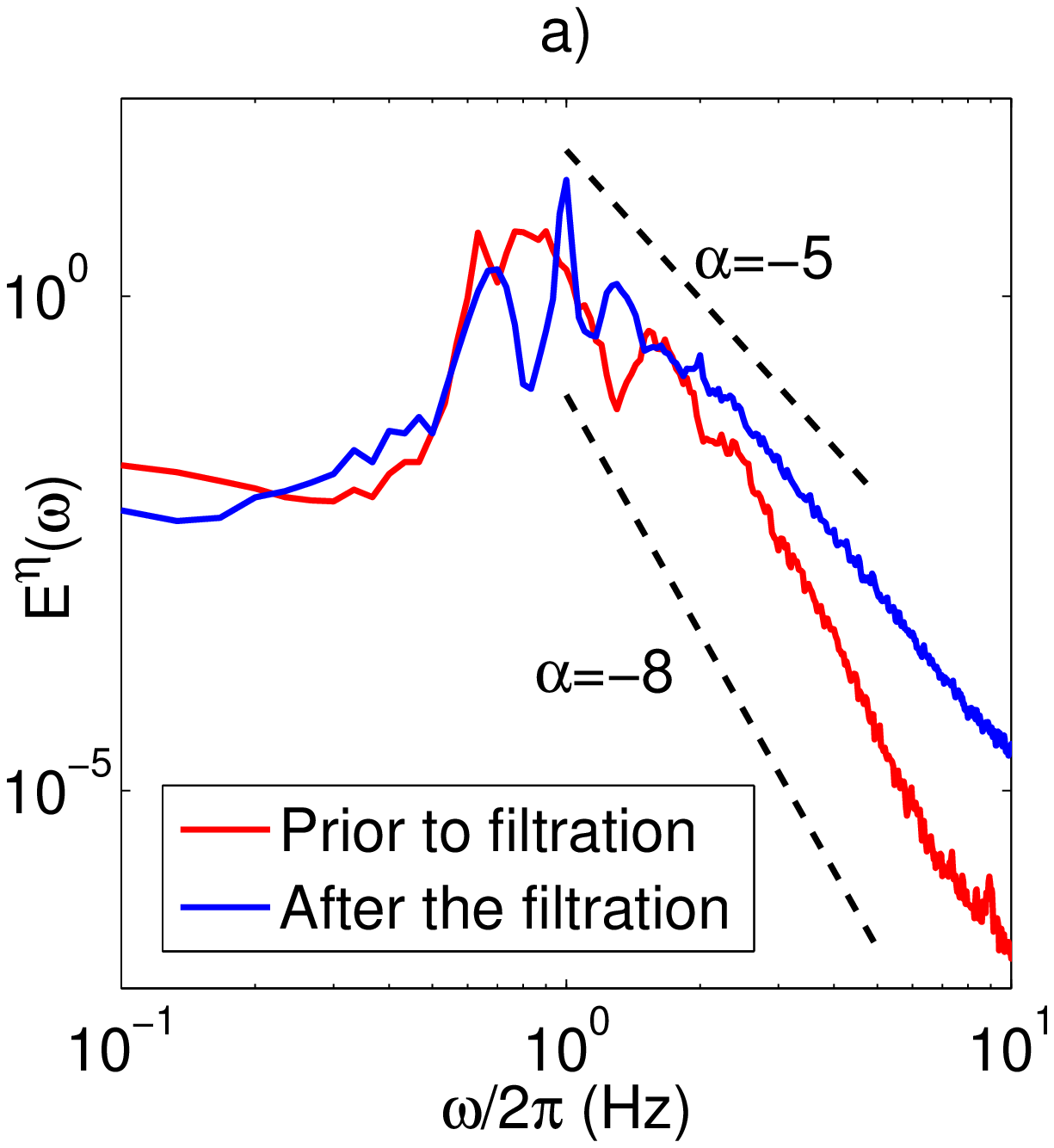}
\includegraphics[width=8cm]{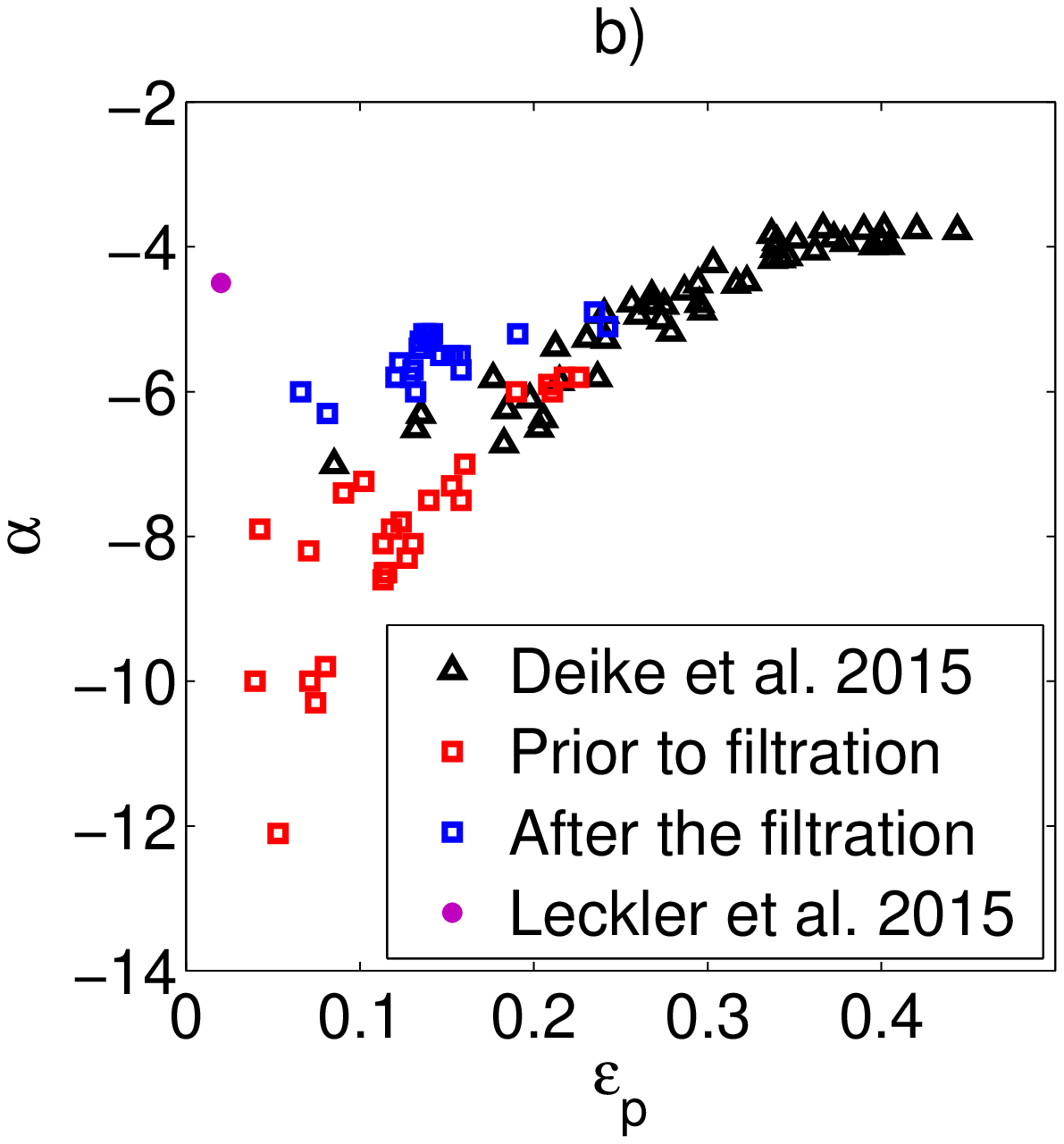}
\caption{(a)Temporal power spectrum $E^\eta(\omega)$ for two experiments with a similar wave steepness. The red curve is obtained without filtration of the free surface while the blue is obtained after a few hours of filtration. The exponent of the spectrum changes from about $-8$ initially to $-5$ for the cleanest water. (b) Distribution of the measured spectral exponent $\alpha$ as function of the typical steepness of the waves $\epsilon_p$ (changed by tuning the magnitude of the forcing). Dark triangles are previous measurements reported by Deike et al. \cite{Deike2015}. The blue and red squares are respectively our measurements with and without filtration. The single purple point is an in-situ measurement of gravity waves in the Black sea \cite{Leckler2015}.}
\label{fig4}
\end{figure}
A quantification of these visual differences is obtained by computing the frequency power spectrum $E^{\eta}(\omega)=\langle\left|\eta(\omega)\right|^2\rangle$ estimated from the capacitive probes. $\eta(\omega)$ is the Fourier transform in time over a given time window and the average $\langle \cdot \rangle$ is a temporal average over successive time windows. Figure~\ref{fig4}(a) shows two power spectra corresponding to two experiments with similar steepness (but slightly distinct forcing intensity) in which data was recorded either before (red curve) or after filtration (blue curve).
As expected by the visualization of the free surface, the power spectrum obtained with a filtered surface shows an exponent close to $-5$ whereas the exponent is about $-8$ for polluted water that corresponds to a much steeper spectrum. Figure~\ref{fig4}(b) shows several measurements of $\alpha$ obtained by fitting a power law in a range of wavevectors where the spectrum is behaving as a power law. The fitting range is usually quite narrow, between $\omega/2\pi=2$~Hz and $\omega/2\pi=5$~Hz at best. The spectral exponent is shown as a function of the typical wave steepness $\epsilon_p$ which is a usual quantification of the intensity of non-linearities. As no direct measurement of the wave steepness is possible with a local measurement, an estimation from the power spectrum is used : $\epsilon_p=2k_p \sigma_\eta$ where $k_p$ is the wave number of the principal peak of the power spectrum (same definition as in \cite{Onorato2009}). A good agreement (within 10\%) with the direct measurement $\epsilon=\sqrt{(\partial\eta/\partial x)^2+(\partial\eta/\partial y)^2}$ has been checked with the spatial reconstruction : $\epsilon_p\approx\sigma_{\epsilon}=\sqrt{\langle\epsilon-\left\langle \epsilon\right\rangle\rangle^2 }$ (where the slope is averaged in time and space to obtain $\langle\epsilon\rangle$). Although both estimates are aimed at quantifying the typical slope, such a good quantitative agreement is most likely a coincidence rather than a general rule. Dark triangles are measurements reported by Deike {\it et al.}~\cite{Deike2015} in a $15\times10$~m$^2$ wave tank (similar size than ours but with a rectangle shape rather than circular). Blue and red squares are our measurements with and without the filtration. A good matching with the measurements of Deike et al. \cite{Deike2015} is observed, showing that Deike's experiments are also most likely impacted by surface pollution. However, even with the filtration, the theoretical exponent $\alpha = -4$ given by the WTT is not reached at low wave steepness. This suggests that the filtration may not be efficient enough and that a residual pollution remains and steepens the spectra. Nonetheless the surface dissipation has clearly being strongly reduced by filtration as the spectral exponent increases when cleaning the water surface. Another possibility is that  finite size effects that may be dominant. Yet, these measurements emphasize that great care must be taken in order to minimize the problem of water pollution that affects the power spectrum in a major way. And this even for a gravity wave experiments where the large scales of the experiment may lead one to believe to be insensitive to surface contamination. Indeed, the maximum of dissipation occurs near $3$~Hz which correspond to a wavelength of about $15$~cm for a linear wave. To obtain one decade in frequency of waves that are unaffected by this pollution, we would need to force waves at $0.1$~Hz with a wavelength of about $150$~m. So even in our large wave tank, we cannot ignore this pollution. This condition can be fulfilled in Nature as shown by the data point (purple in Fig.~\ref{fig4}(b)) extracted from Black Sea data by Leckler~{\it et al.}. The nonlinearity is much lower than in the experiment but the value of the exponent is very close to the theoretical value of $-4$. The larger scale separation may be responsible for this better agreement between observation and theory. The surface pollution may also have been advected away by the wind that was blowing from the shore so that the water surface could be very clean as well.

\subsubsection{Full wavevector-frequency spectrum}

\begin{figure}[!htb]
\includegraphics[width=8cm]{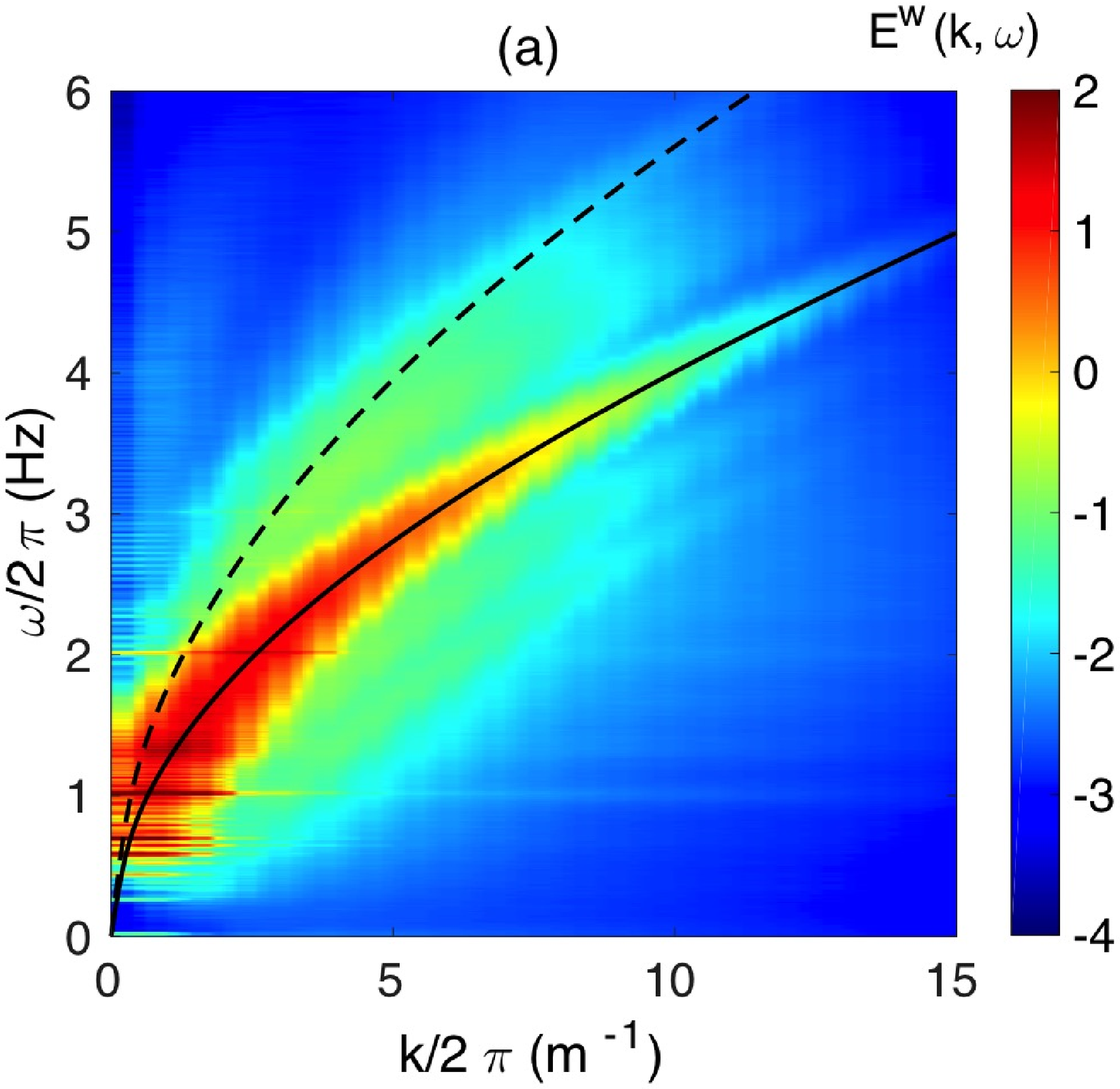}
\includegraphics[width=8cm]{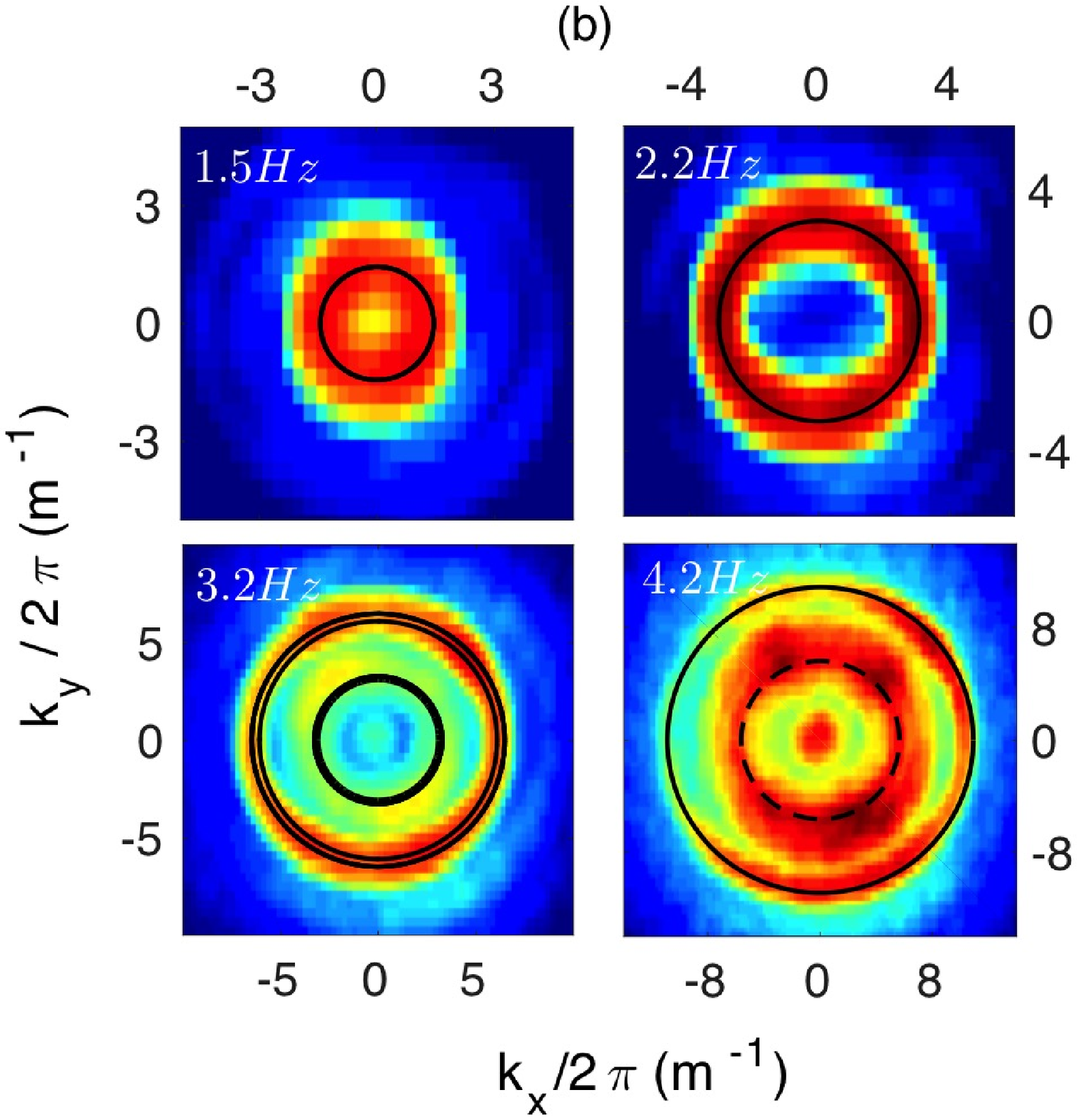}
\caption{(a) Space time Fourier power spectrum of the vertical velocity of the waves $E^w(k,\omega)$ for a typical experiment where the mean wave steepness is $\epsilon=0.11$. The spectrum $E^w(\mathbf k,\omega)$ has been integrated over directions of $\mathbf k$. (b) Four  cross sections of the complete spectrum $E^w(\mathbf{k},\omega)$ at given frequencies (values indicated on the image). In both cases, the solid black line is the deep water linear dispersion relation $\omega=\sqrt{gk}$ and the dashed line is its second harmonic. Energy is color-coded in log-scale.}
\label{fig5}
\end{figure}
In the following, only data recorded in a single experiment with a typical wave steepness of $\epsilon_p=0.11$ will be presented. We use the Stereo-PIV technique to measure the velocity of the water surface resolved in space and time. Figure~\ref{fig5} shows the full power spectrum of the vertical velocity $E^w(\mathbf{k},\omega)=\langle\left|w(\mathbf{k},\omega)\right|^2\rangle$. Figure \ref{fig5}(a) shows the power spectrum $E^w(k,\omega)$ integrated over the direction of $\mathbf{k}$. Energy is mainly concentrated along the linear dispersion relation of deep water gravity waves $\omega=\sqrt{gk}$ (solid line). However for higher frequencies, a significant part of the energy lies away from the linear dispersion relation, in particular along the second harmonic $\omega=\sqrt{2gk}$ (dashed line). This point is clearly seen in Fig.~\ref{fig5}(b) that shows cross-sections in the $(k_x,k_y)$ plane of the full spectrum $E^w(\mathbf{k},\omega)$ at four given frequencies $1.5$, $2.2$, $3.2$ and $4.2$~Hz. In addition to the observation of a quasi-isotropic system, we notice that at $4.2$~Hz a major part of the energy is concentrated on the second harmonic which seems to indicate a strong level of non-linearities. This observation is potentially related to the limited spatial resolution of the measurement. As we study large values of $\omega$, harmonic waves have a lower $k$ than linear waves. At the highest frequencies, the size of the correlation box in the Stereo-PIV measurement does not allow us to measure linear waves, which amplitude is reduced by filtering, while harmonics remain observable because of their larger wavelengths at the same frequency. Space-time spectra were also observed recently in experiments\cite{Taklo1,Taklo2} by using arrays of local probes and the repeatability of the generation to construct a synthetic arrays with sufficient resolution. They observed also energy out of the dispersion relation with presence of harmonics. The excitation was rather narrowband and interpreted in the framework of the nonlinear Schrödinger equation.
 
\begin{figure}[!htb]
\includegraphics[width=8cm]{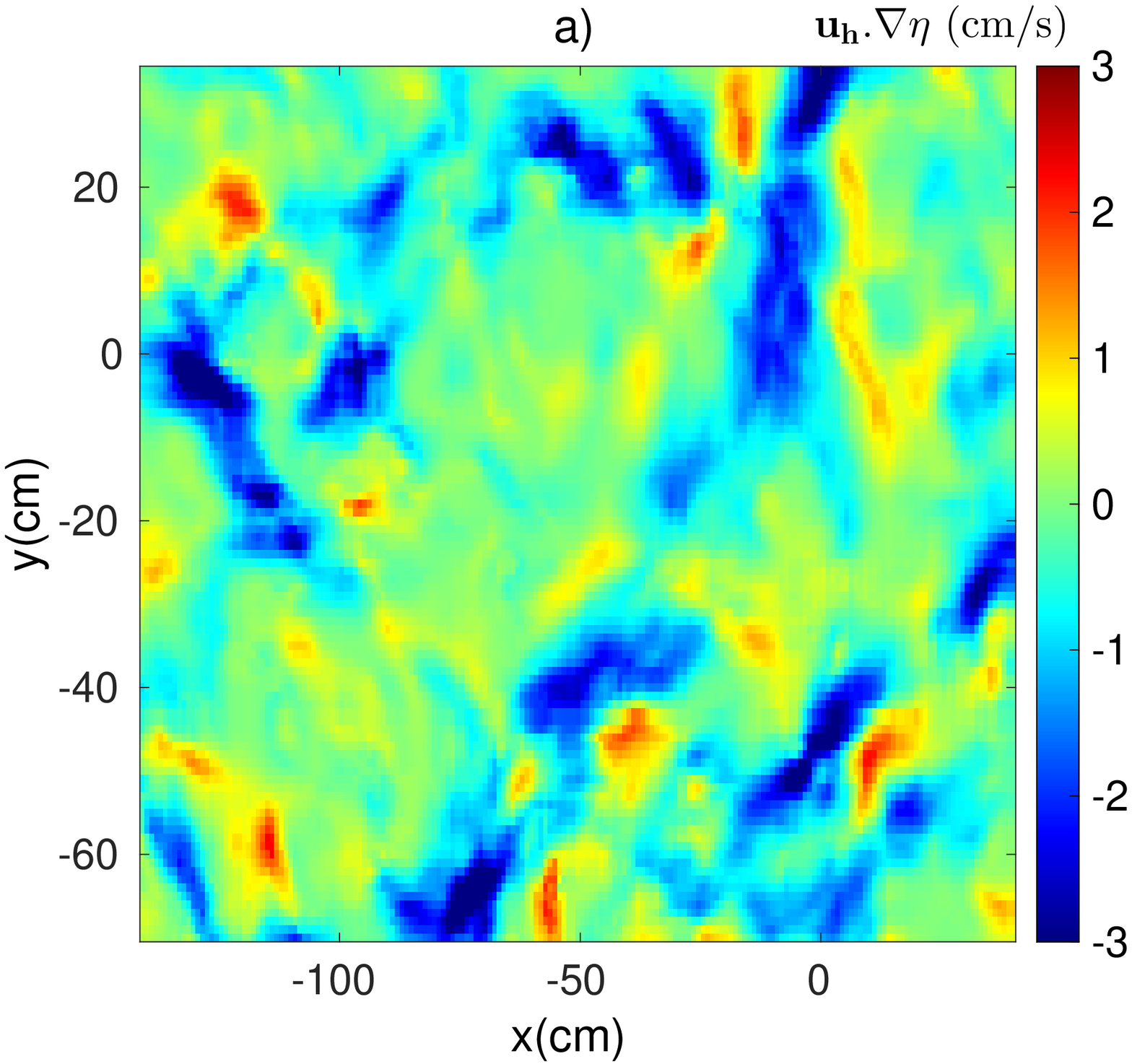}
\includegraphics[width=8cm]{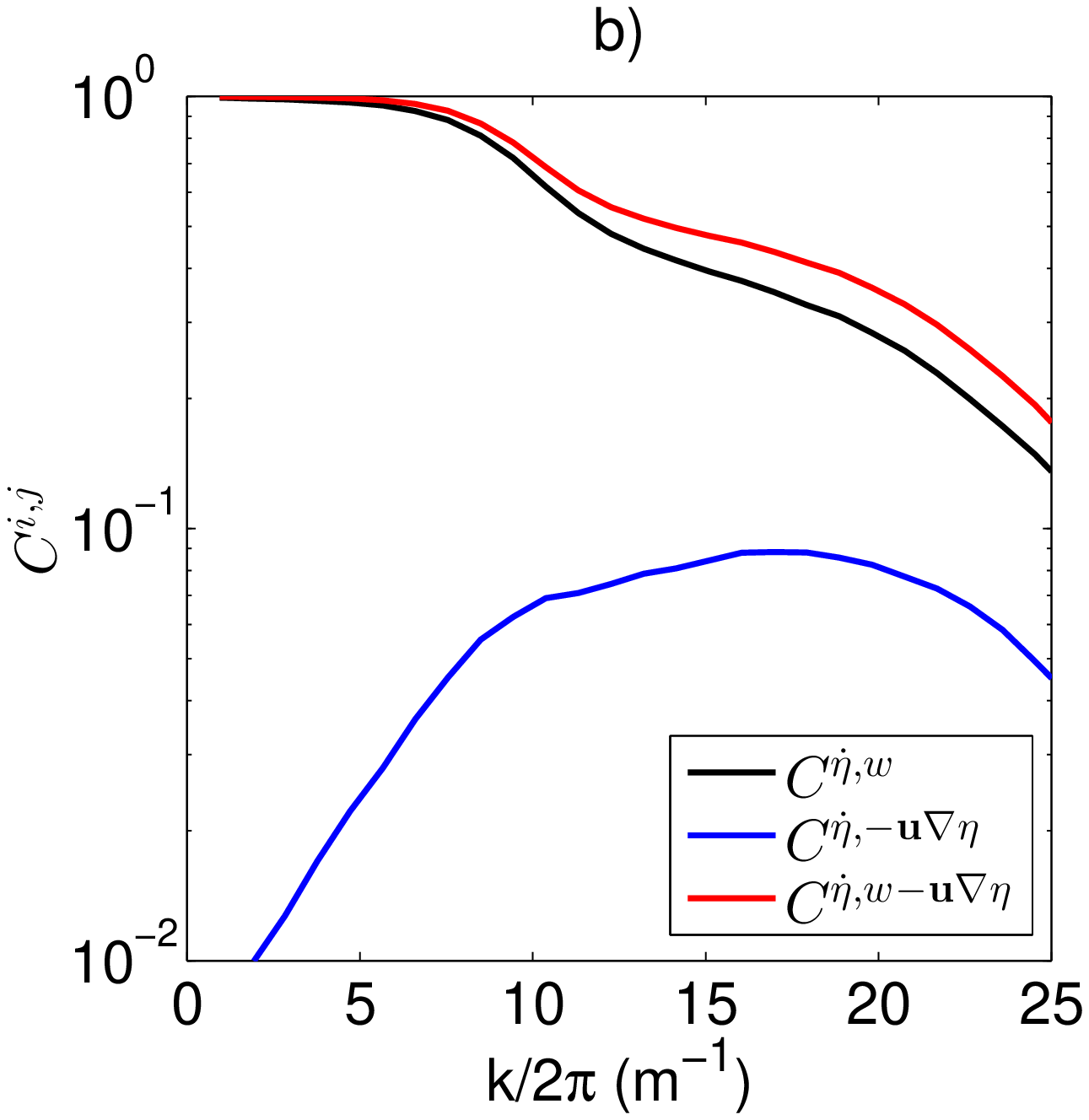}
\caption{(a) Snapshot of $\mathbf{u_h}\cdot\nabla \eta$, the non-linear component of the vertical velocity field $w=\frac{\partial \eta}{\partial t}+\mathbf{u_h}\cdot\nabla \eta$. (b) Coherence of the different components of the velocity relation (\ref{equvel}) as a function of $k$. Assuming isotropy, the coherence is averaged over eight directions of $\mathbf k$ regularly spread over a circle. The black curve is the coherence $C^{\dot{\eta},w}$ between the vertical geometric velocity $\dot{\eta}$ and $w$. The blue curve is the coherence $C^{\dot{\eta},-\mathbf{u_h}\cdot\nabla \eta}$ between $\dot{\eta}$ and $-\mathbf{u_h}.\nabla \eta$. The red curve is the coherence $C^{\dot{\eta},w-\mathbf{u_h}}$ between $\dot{\eta}$ and $w-\mathbf{u_h}\cdot\nabla \eta$ (see (\ref{eq:coh}) for definitions).}
\label{fig5b}
\end{figure}
In order to perform a quantitative analysis of the strength of the non-linearities, we compute the non-linear term $\mathbf{u_h}\cdot\nabla \eta$ in the the kinematic surface condition (a snapshot is shown in Fig.~\ref{fig5b}(a)) : 
\begin{equation}
\frac{\partial \eta}{\partial t}=w-\mathbf{u_h}\cdot\nabla \eta
\label{equvel}
\end{equation}
where $\mathbf{u_h}=(u,v)$ is the horizontal surface velocity. Figure~\ref{fig5b}(b) shows a computation of the spectral coherence $C$ between the different terms of equation (\ref{equvel}) defined as:
\begin{equation}
\begin{array}{c}
C^{\dot{\eta},w}(\mathbf k)=\frac{\left|\left\langle  \dot{\eta}^*(\mathbf k,t)w(\mathbf k,t)\right\rangle\right|}{\sqrt{\left\langle \left| \dot{\eta}\right|^2\right\rangle \left\langle \left| (w-\mathbf{u_h}\cdot\nabla \eta) \right|^2\right\rangle}}\\
C^{\dot{\eta},-\mathbf{u_h}\cdot\nabla \eta}(\mathbf k)=\frac{\left|\left\langle  \dot{\eta}^*(\mathbf k,t)(-\mathbf{u_h}\cdot\nabla \eta)(\mathbf k,t)\right\rangle\right|}{\sqrt{\left\langle \left| \dot{\eta}\right|^2\right\rangle \left\langle \left| (w-\mathbf{u_h}\cdot\nabla \eta) \right|^2\right\rangle}}\\
C^{\dot{\eta},w-\mathbf{u_h}\cdot\nabla \eta}(\mathbf k)=\frac{\left|\left\langle  \dot{\eta}^*(\mathbf k,t)(w-\mathbf{u_h}\cdot\nabla \eta)(\mathbf k,t)\right\rangle\right|}{\sqrt{\left\langle \left| \dot{\eta}\right|^2\right\rangle \left\langle \left| (w-\mathbf{u_h}\cdot\nabla \eta) \right|^2\right\rangle}}
\end{array}
\label{eq:coh}
\end{equation}
where the average $\langle \cdot \rangle$ is an average over time and $\dot{\eta}=\frac{\partial \eta}{\partial t}$.

Using our choice of a common normalization of the coherence allows us to compare directly the three coherence estimators along the dimension $k$. The coherence $C^{\dot{\eta},w-\mathbf{u_h}.\nabla \eta}$ (red line) should be equal to one in a perfect measurement. However, the inevitable presence of noise decreases this level at high $k$ (where the signal over noise ratio is getting lower). We observe that the coherence fraction of the non-linear term increases with $k$ up to a maximum around $10$\% before dropping due to the noise. The proximity of the black and red curve confirms the weakly-nonlinear character of our system which is thus expected to be in the range of validity of the WTT, as far as the strength of non-linearity is concerned.

\section{Investigation of 3-wave interactions}
\label{3w}

\subsubsection{Theoretical analysis}
\begin{figure}[!htbp]
\includegraphics[width=11cm]{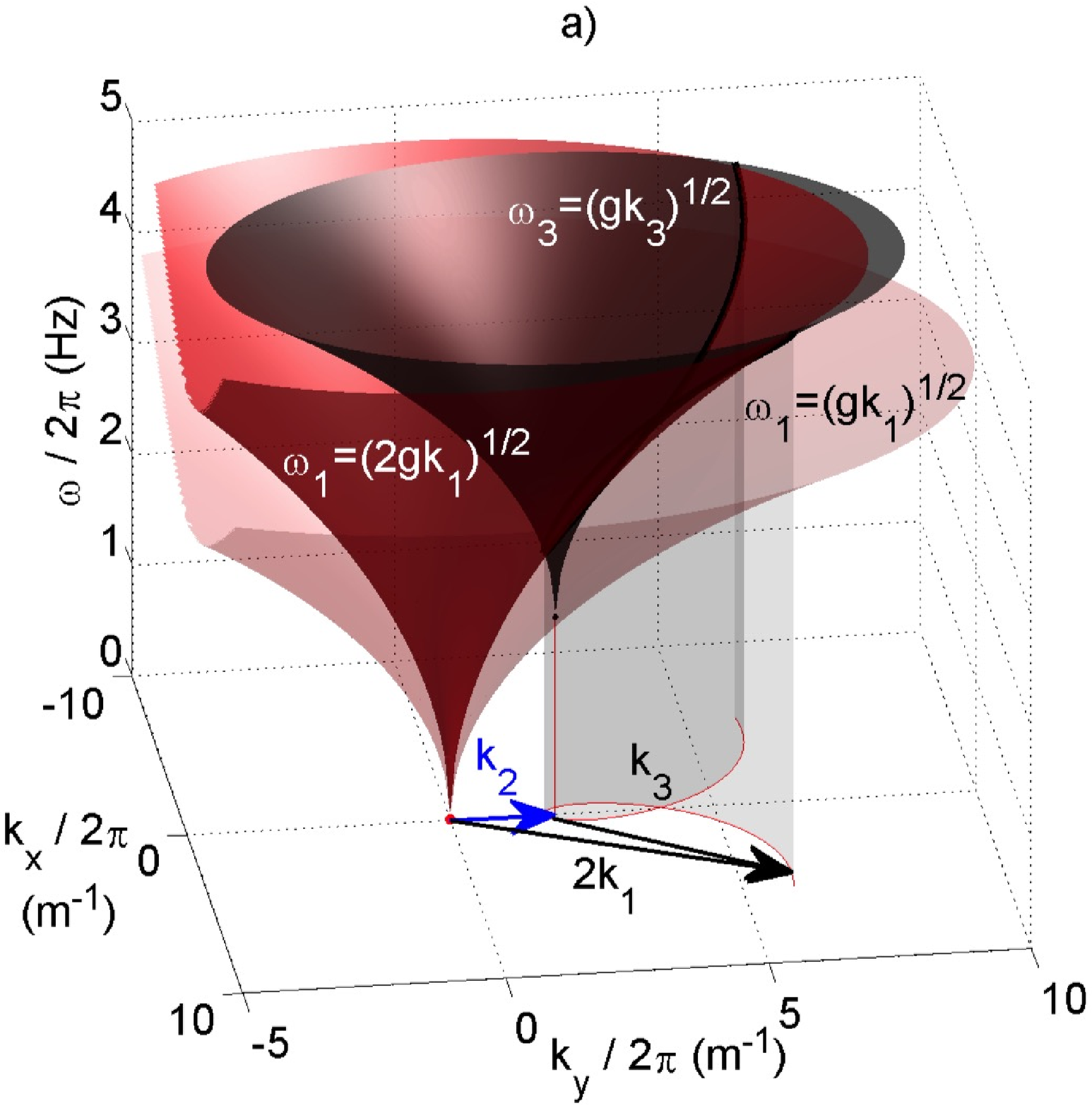}
\includegraphics[width=8cm]{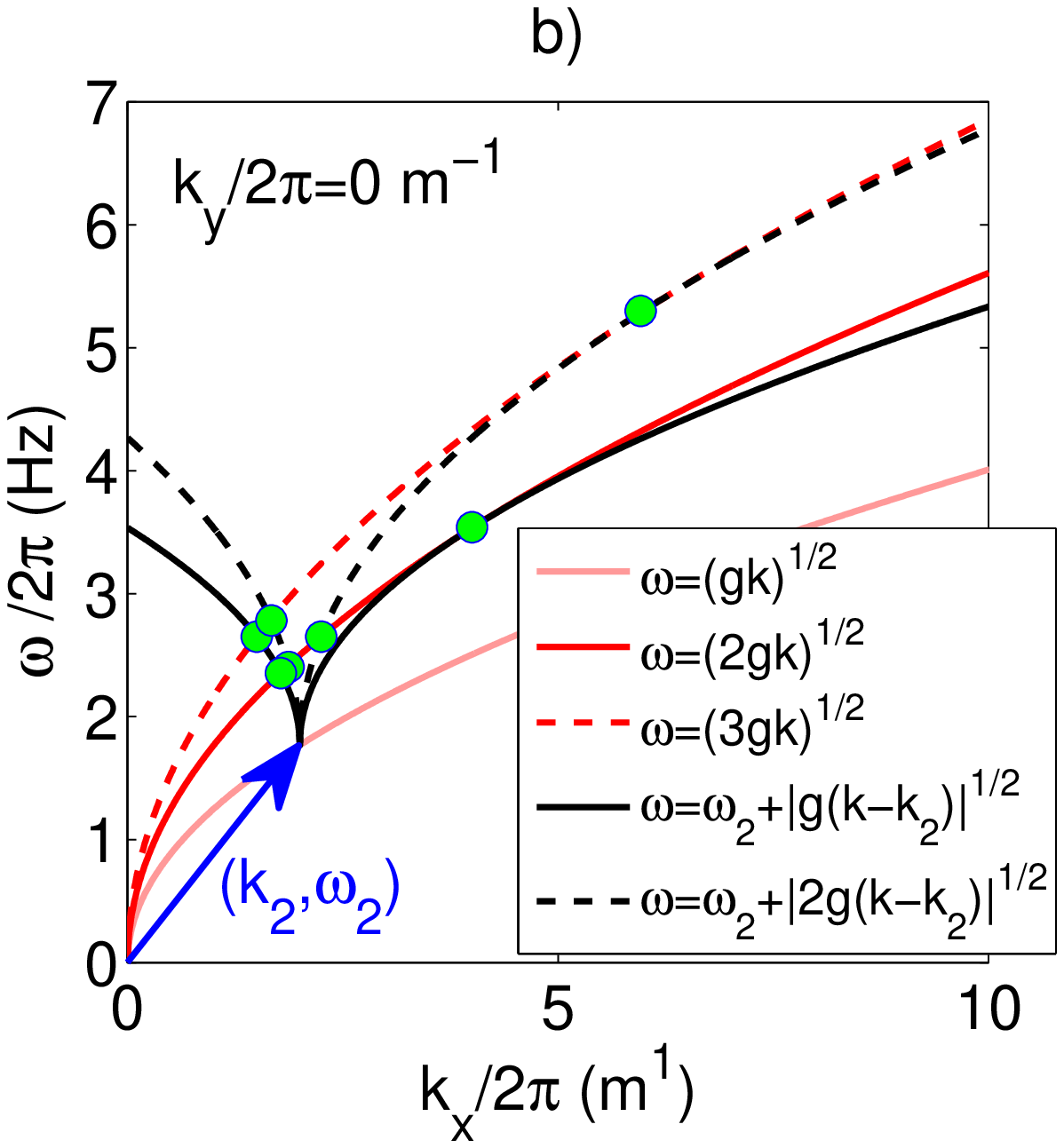}
\includegraphics[width=8cm]{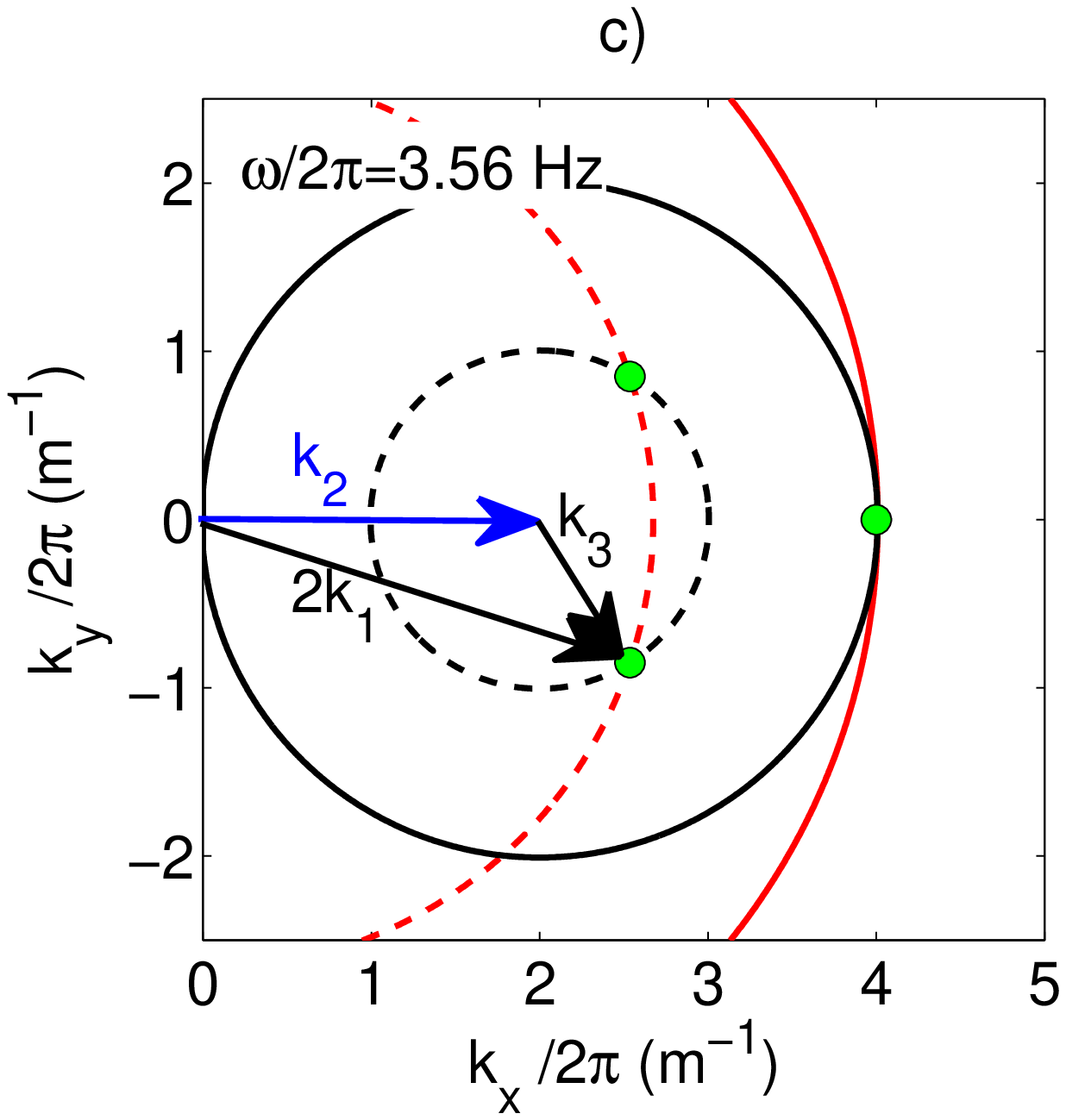}
\caption{(a) 3D Theoretical exact geometrical solution for 3-wave resonant interactions. For simplicity, only the positive frequencies are represented. A first linear dispersion relation $\omega_1(\mathbf k_1)$ is shown in light pink. We pick a value of the pair $(\mathbf k_2,\omega_2)$ shown with the blue arrow. We then plot a second linear dispersion relation in black starting from the point $(\mathbf k_2,\omega_2)$. This black surface is thus $\omega_2+\omega_3(\mathbf k_3=\mathbf k-\mathbf k_2)$. Any intersection of both surfaces corresponds to resonance $\omega_1=\omega_2+\omega_3$. As it is well known, we observe no intersection between the pink and black curves, as no 3-wave resonances are geometrically allowed for wave following the linear dispersion relation $\omega=\sqrt{gk}$. The red surface is the second harmonic $\omega_1=\sqrt{2gk_1}$. The intersection of the red and the black surface indicates the presence of resonant solutions when the wave labeled 1 is a second harmonic of the linear dispersion relation and the waves 2 and 3 are linear waves. These are usually known as bound harmonics where $\mathbf k_2+\mathbf k_3=2\mathbf k_1$. (b) Cross-section of the 3D representation for $k_y/2\pi=0$~m$^{-1}$. The third harmonic $\omega=\sqrt{3gk}$ is added as the red dashed line and the second harmonic $\omega_3=\sqrt{2gk_3}$ is added as the black dashed line starting from the point $(k_2,\omega_2)$. Exact solutions are indicated by the green dots. Note that in the vicinity of some solutions (large values of $k_x$ at the right of the figure), the two dispersion relations remain close for a large range of frequencies, suggesting the possibility of quasi-resonance (see text). (c) Cross-section of (a) for a given frequency $\omega/2\pi=3.56Hz$. Exact solutions are plotted with green dots.}
\label{fig6}
\end{figure}

Both temporal and spatio-temporal observed power spectra are inconsistent with the predictions of the WTT possibly due to additional dissipation or finite size effects as discussed above. Yet, if the theory is applicable to our experiment, resonant interactions should exist and be the main mechanism to transfer energy. 
We start our analysis by investigating $3$-waves interactions. Strictly resonant theoretical solutions can be found geometrically using the linear dispersion relation. Figure~\ref{fig6}(a) shows the ensemble of solutions for a given pair $(\mathbf{k_2},\omega_2)$.
As expected, the dispersion relation of $\omega_1$ (in pink surface) does not intersect with the linear dispersion of $\omega_2+\omega_3$ (black surface). As it is well known for pure gravity waves, this means that no exact $3$-waves solution exist for freely propagating waves following the linear dispersion relation. However, the analysis of the power spectrum showed that up to $10\%$ of the energy lies in the second harmonic and some energy is scattered around the dispersion relation. In the following we consider the correlations between Fourier modes that may not follow the linear dispersion relation but that are resonant nonetheless, i.e. that follow the 3-wave resonance relations (\ref{eq3w}). Some of these modes are typically bound waves that result from the quadratic interaction of two free waves.

We start by including the second harmonic to look for resonant solutions as the red surface $\omega_1=\sqrt{2gk_1}$ in Fig.~\ref{fig6}(a). One observes the existence of an intersection with the black surface $\omega_2+\omega_3$, suggesting the possibility of resonant interactions involving waves that do not necessarily fulfill the linear dispersion relation. Figures~\ref{fig6}(b) and (c) show two cross-sections of the Fig.~\ref{fig6}(a) at given plans $k_y/2\pi=0$~m$^{-1}$ and $\omega/2\pi=3.56$~Hz (resp.) for a better view of the solutions (green dots). Among these solutions, one is the special case of a wave interacting with itself to form its own harmonic : $\mathbf k_2=\mathbf k_3=\mathbf k_0$ so that $\mathbf k_2+\mathbf k_3=\mathbf k_1=2\mathbf k_0$  (and similar equations for $\omega$). 

We also note that in vicinity of this bound wave solution, the two dispersion curves (black and red curves) remain very close to each other. This may allow quasi-resonant interactions :
\begin{equation}
\mathbf{k_2}+\mathbf{k_3}=\mathbf{k_1} \quad\quad \omega_2+\omega_3=\omega_1+\delta \omega
\label{eq3wdk}
\end{equation}
where $\delta \omega$ is the detuning that will allows resonance if the width $\Delta \Omega$ of the power spectrum is larger ($\Delta \Omega>\delta \omega$). From Fig.~\ref{fig5}, it can be seen that $\Delta \Omega$ varies significantly across the Fourier space but it takes typical values of order a few tenth of Hz while Fig.~\ref{fig6}(b) shows that the distance between the continuous red and black curves remains extremely small across a wide interval around their intersection (at the center of the figure). From this observation we may expect quasi-resonant interactions over an interval of frequencies of width about 1~Hz. However, $\Delta \Omega$ is far too small to generate resonant interactions between the two linear branches (pink and black line, separated by about 1 Hz), in contrast to what has been observed in gravity-capillary waves turbulence~\cite{Aubourg2015,Aubourg2016}.

The second harmonic $\omega_1=\sqrt{2gk_1}$ and the third harmonic $\omega_1=\sqrt{3gk_1}$ are also shown in Fig.~\ref{fig6}(b) with the dashed black line and the dashed red line respectively. They show similar features so that one could expect quasi-resonances in between second and third harmonics.

\subsubsection{Correlations}
\begin{figure}[!htbp]
\includegraphics[width=8cm]{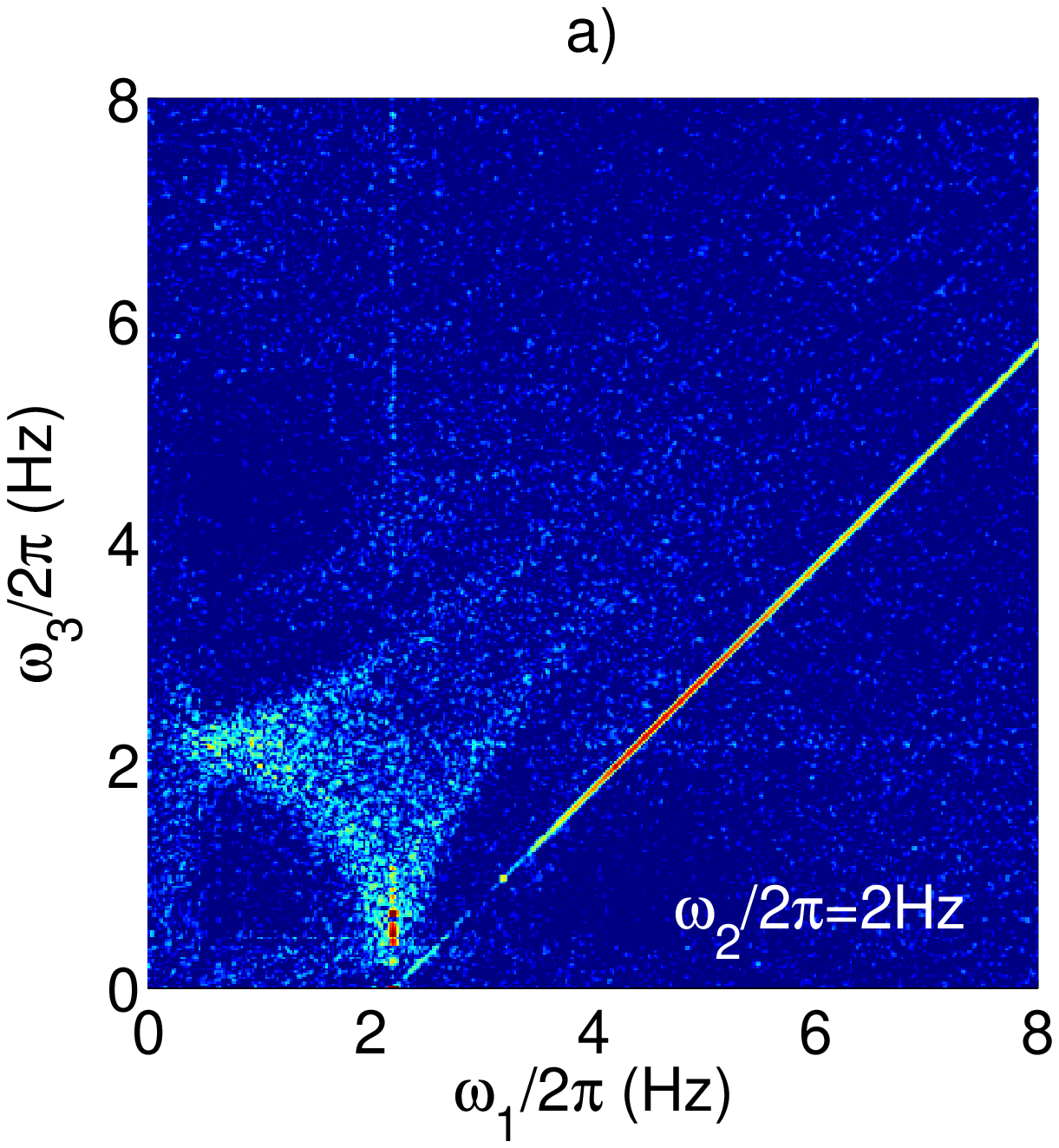}
\includegraphics[width=8cm]{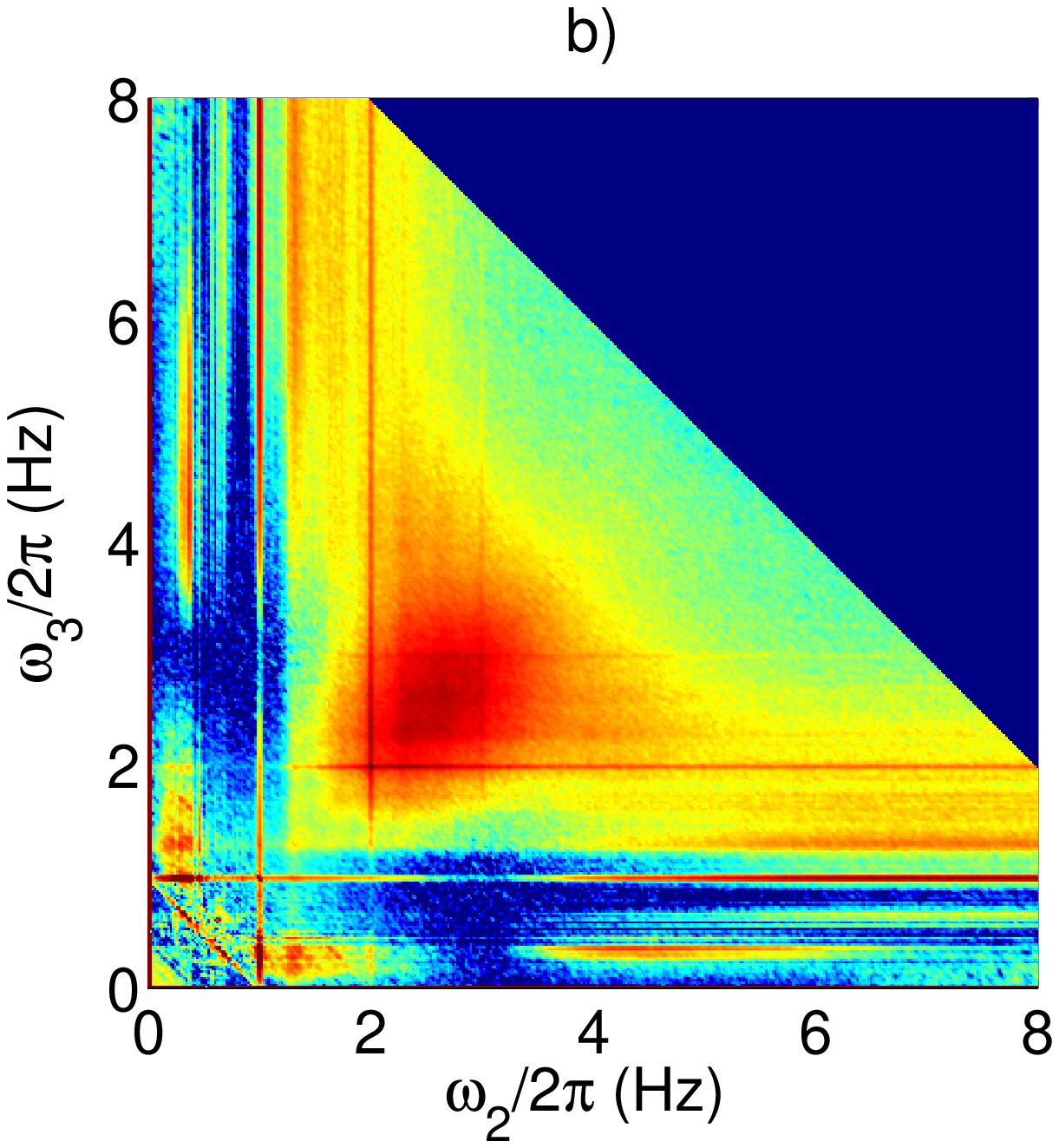}\\
\includegraphics[width=8cm]{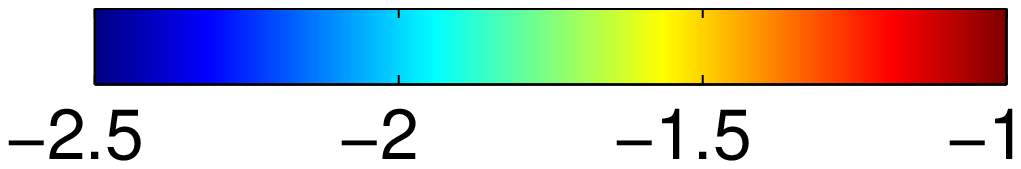}
\caption{a) Normalized third order correlation $C_3^{\omega}(\omega_1,\omega_2,\omega_3)$ for a given wave $\omega_2/2\pi=2$~Hz. A high concentration of correlations is present along the resonant line $\omega_1=\omega_2+\omega_3$, suggesting the presence of 3-wave interactions. b) Bicoherence $B_3^{\omega}(\omega_2,\omega_3)$. A red spot is visible around $\omega_2=\omega_3=2\pi\times2.5$~Hz.}
\label{fig7}
\end{figure}
$3$-wave resonant interactions are investigated using third order correlations. We start the analysis in frequency space using the third order correlation of the vertical velocity. 
\begin{equation}
c^3_{\omega}(\omega_1,\omega_2,\omega_3)=\left\langle w^*(\omega_1)w(\omega_2)w(\omega_3)\right\rangle
\end{equation}
in order to check the presence of resonant frequency triads satisfying $\omega_1=\omega_2+\omega_3$. Figure~\ref{fig7}(a) shows this correlation estimator for a given frequency $\omega_2/2\pi=2$~Hz and normalized so that to ensure $0\leq C^3_\omega \leq 1$ \cite{Collis1998}:
\begin{equation}
C^3_{\omega}(\omega_1,\omega_2,\omega_3)=\frac{c^3_{\omega}(\omega_1,\omega_2,\omega_3)}{\sqrt{\left\langle \left|w(\omega_2)w(\omega_3)\right|^2\right\rangle\left\langle \left|w(\omega_{1})\right|^2\right\rangle}}
\label{C3w}
\end{equation}

The resonant line $\omega_1=\omega_2+\omega_3$ is clearly observed above the statistical noise (which level is about $10^{-2.5}$) indicating the presence of 3-wave resonances in the system. Figure~\ref{fig7}(b) shows the bicoherence: 
\begin{equation}
B^3_{\omega}(\omega_2,\omega_3)=C^3_{\omega}(\omega_2+\omega_3,\omega_2,\omega_3)
\end{equation}
which corresponds to the extraction of the resonant line of $C^3_{\omega}$. A wide spot of strong correlations (about $25\%$) appears when $\omega_2\gtrsim 2.5$~Hz and $\omega_3\gtrsim 2.5$~Hz as well. The high correlation values for similar values of $\omega_2$ and $\omega_3$ indicates that coupling with harmonic waves may be present close to the bound waves solutions where $\omega_2$ is equal to $\omega_3$.

\begin{figure}[!htb]
\includegraphics[width=8cm]{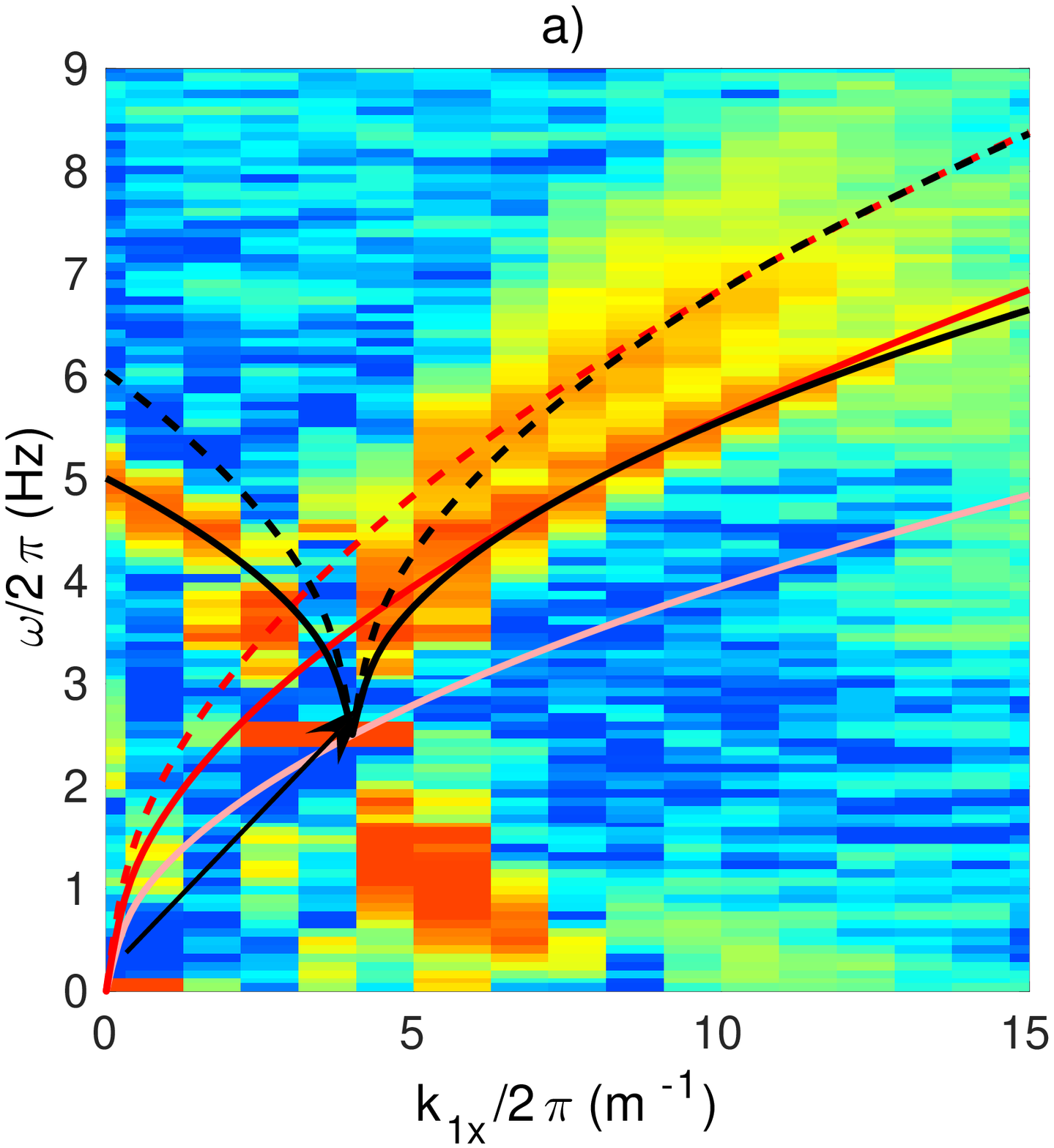}
\includegraphics[width=8cm]{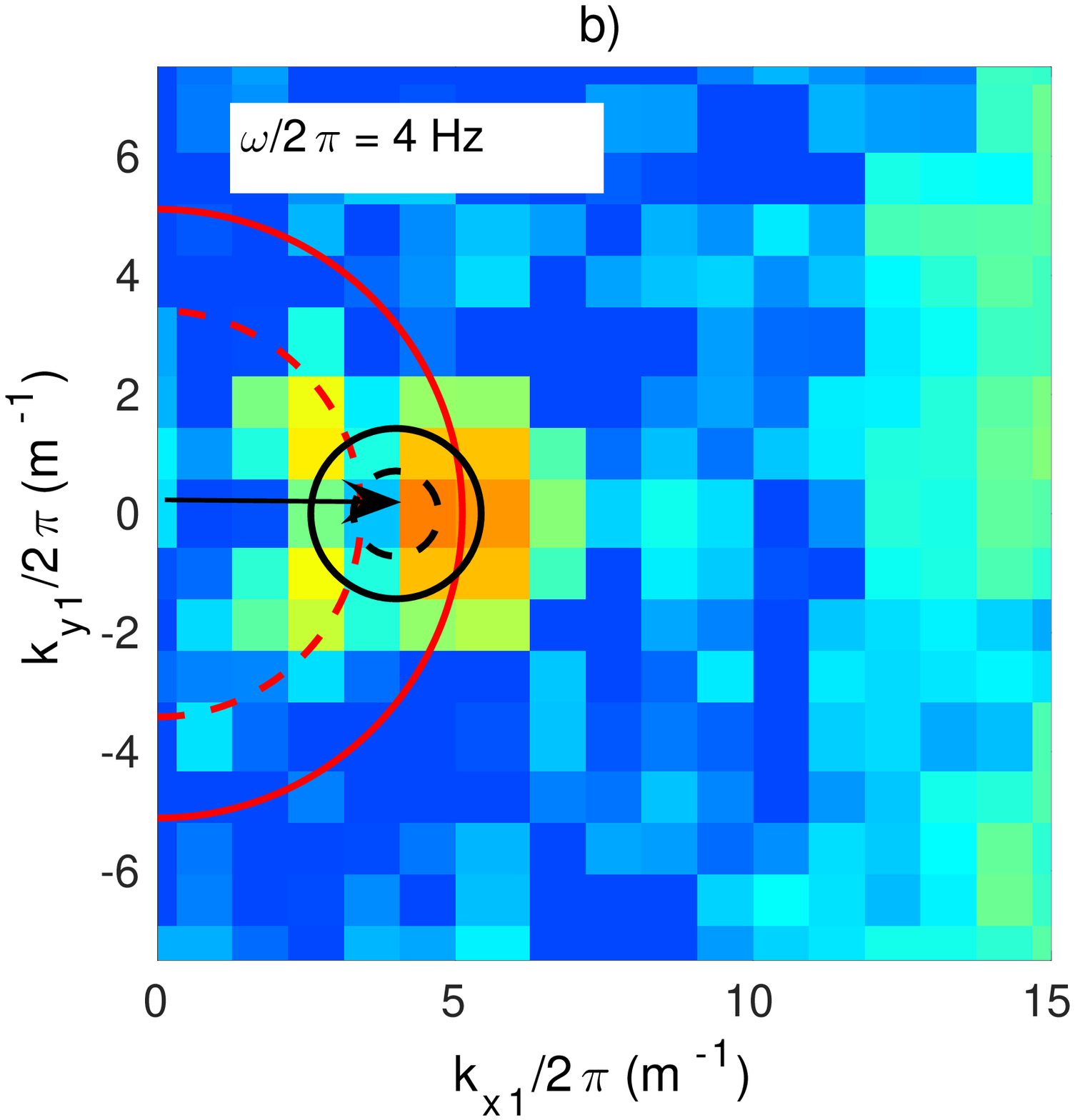}
\includegraphics[width=8cm]{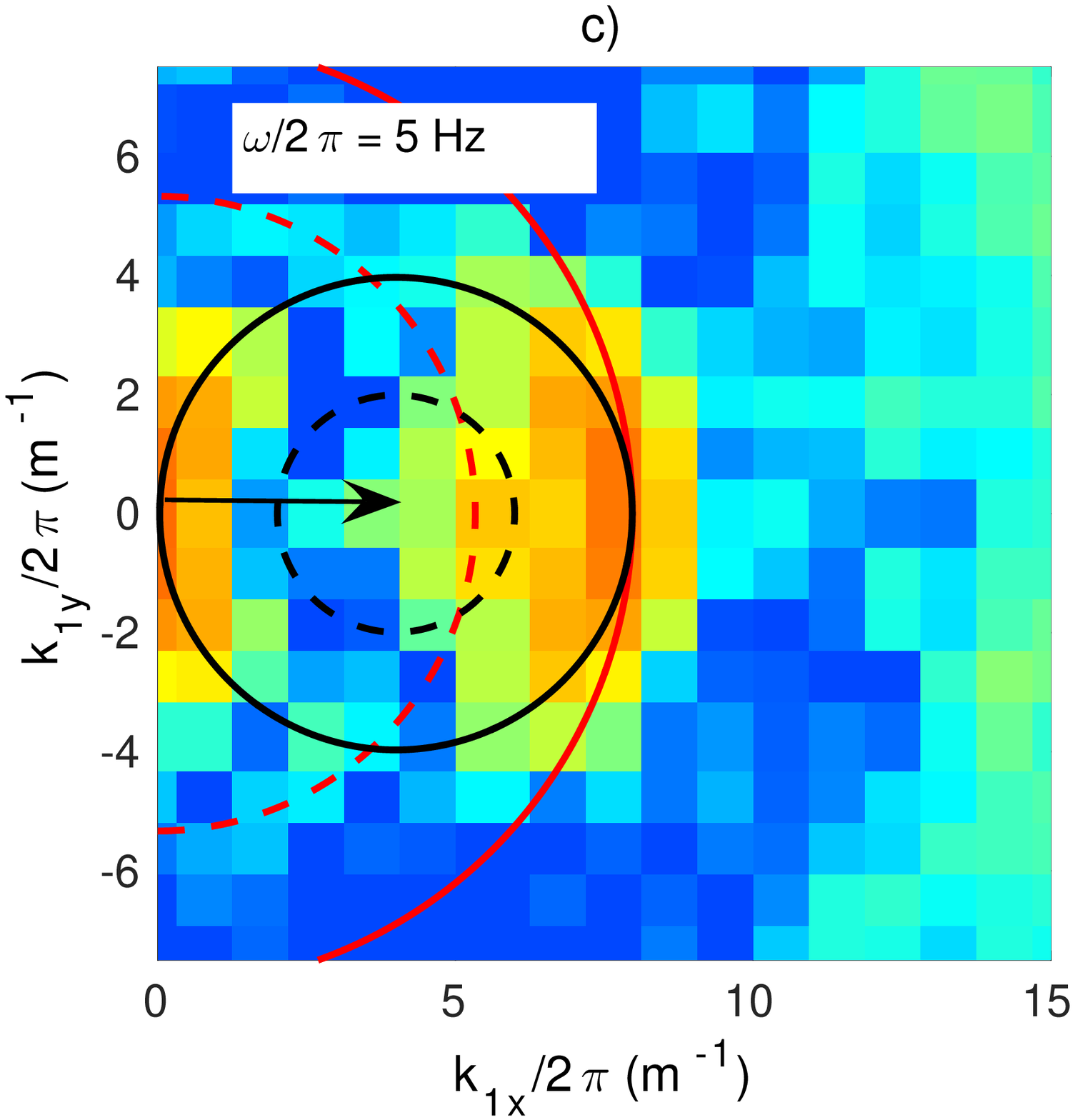}
\includegraphics[width=8cm]{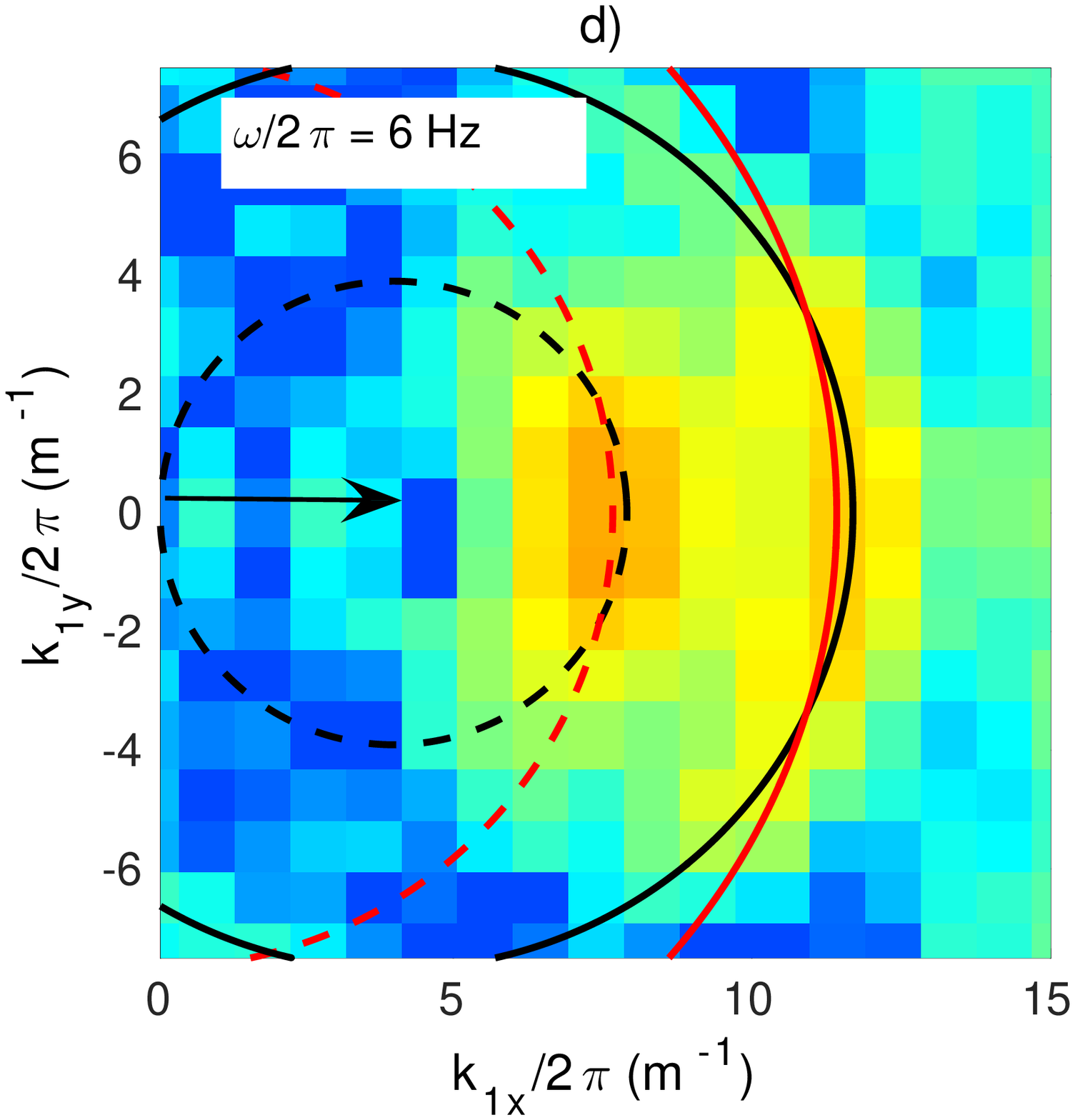}
\includegraphics[width=8cm]{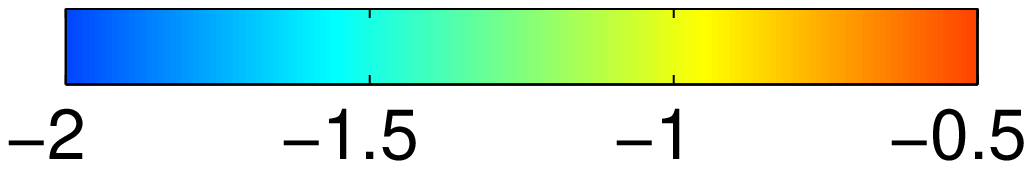}
\caption{3D Bicoherence $B^{k,\omega}_3(\mathbf{k_2},\mathbf{k_3},\omega_2,\omega_3)$ for a given wave $(\mathbf{k_2},\omega_2)$ (with $|k_2|/2\pi=4$~m$^{-1}$). In order to allow a direct comparison, the axes are the same than those of figure~\ref{fig6} and show $\mathbf k_1=\mathbf k_2+\mathbf k_3$ and $\omega_1=\omega_2+\omega_3$. The $x$ axis is given by the direction of $\mathbf k_2$. The black arrows show the wave $(\mathbf{k_2},\omega_2)$. (a) Cross-section in the plane $(k_x,k_y=0,\omega)$. The red and black curves are similar to ones shown in fig \ref{fig6}. Exact resonances with harmonics are at the intersection of a black curve and a red one. A high correlation intensity is observed in these regions, suggesting the occurrence of interaction between linear waves and harmonics. (b), (c) and (d) are three cross-sections in the planes corresponding to given values of $\omega/2\pi=4$, $5$ and $6$~Hz. An angular dispersion of the interactions of about $90^{\circ}$ is observed.}
\label{fig8}
\end{figure}

To obtain a more precise view of the coupling we have to extend the analysis in the wavenumber and frequency domain in order to distinguish the different branches (linear and harmonics). This can be done with the space and time bicoherence:
\begin{equation}
B^3_{\mathbf{k},\omega}(\mathbf{k_2},\mathbf{k_3},\omega_2,\omega_3)=\frac{\left\langle w^*(\mathbf{k_2},\omega_2)w(\mathbf{k_3},\omega_3)w(\mathbf{k_2}+\mathbf{k_3},\omega_2+\omega_3)\right\rangle}{\sqrt{\left\langle \left|w(\mathbf{k_2},\omega_2)w(\mathbf{k_3},\omega_3)\right|^2\right\rangle\left\langle \left|w(\mathbf{k_2}+\mathbf{k_3},\omega_{2}+\omega_{3})\right|^2\right\rangle}}
\label{C3kw}
\end{equation}
The large dimensionality of parameter space imposes to choose the value of some components to be more easily understandable. To be in the same representation as the theoretical solution presented in Fig.~\ref{fig6}, we impose a wave $(\mathbf{k_2},\omega_2)$ on the linear dispersion relation. In order to improve the statistical convergence, we average the bicoherence over eight directions of the wavevector $\mathbf k_2$ (isotropically spread in direction). Figure~\ref{fig8} shows four cross-sections of $B^3_{\mathbf{k},\omega}$ for $\left|\mathbf{k_2}\right|/2\pi=4$~m$^{-1}$. 

The cross-section corresponding to the unidirectional solutions ($k_y=0$) is plotted in Fig.\ref{fig8}(a) in the same representation as Fig.~\ref{fig6}(b) . Solid and dotted lines are representing the linear dispersion relations and the second harmonic with the same colorcoding as in Fig.\ref{fig6}. The correlation presents several areas of high intensity above $10\%$. Two of them are located near the exact bound waves solution. We note also that these high correlation regions show a strong spreading lying on the region where the two dispersion curves (linear and harmonic) remain close. This confirms the presence of quasi-resonant interactions in the system which are at the same order of magnitude than the exact bound solution. It is noticeable that the exact bound waves solution does not display any specificity as compared to the approximate resonances. As expected, there is no quasi-resonant interaction between the two linear branches because their separation is stronger than the spectrum width $\Delta k/2\pi\approx 0.5$~m$^{-1}$. 

Figures \ref{fig8} (b), (c) and (d) show three cross-sections at a given frequencies. Again, we observe that high correlation regions are located near the intersections of black and red curves (exact resonances). As seen before in the case of the unidirectional cross-section, a spread of the high correlations is present due to quasi-resonant interactions. This leads to a uniform area of correlation that covers an angular distribution of about $90^{\circ}$ around the direction of $\mathbf k_2$. Some other areas of high correlation can be seen, in particular for negative $\mathbf{k_3}$ (solid black curve) or for negative frequencies $\omega_3$ (in Fig.\ref{fig8}(a), correlations around $(0-2)$~Hz). These results are puzzling since no exact or quasi-resonant solutions can be found to explain their presence. Moreover, the Fourier component at $(\mathbf k_1,\omega_1)$ has extremely weak energy in the power spectrum. We suspect an effect of coupling due to partially standing waves. Indeed due to the finite size of the container all waves are not fully freely propagating waves. Part of the energy is due to standing waves made of the superposition of waves at several wavevectors (but the same frequency) that are phase matched. This could cause additional spurious resonances among seemingly non resonant waves. However no actual proof can be proposed at this time.

\subsubsection{Field data from the Black Sea}

\begin{figure}[!htb]
\includegraphics[width=8cm]{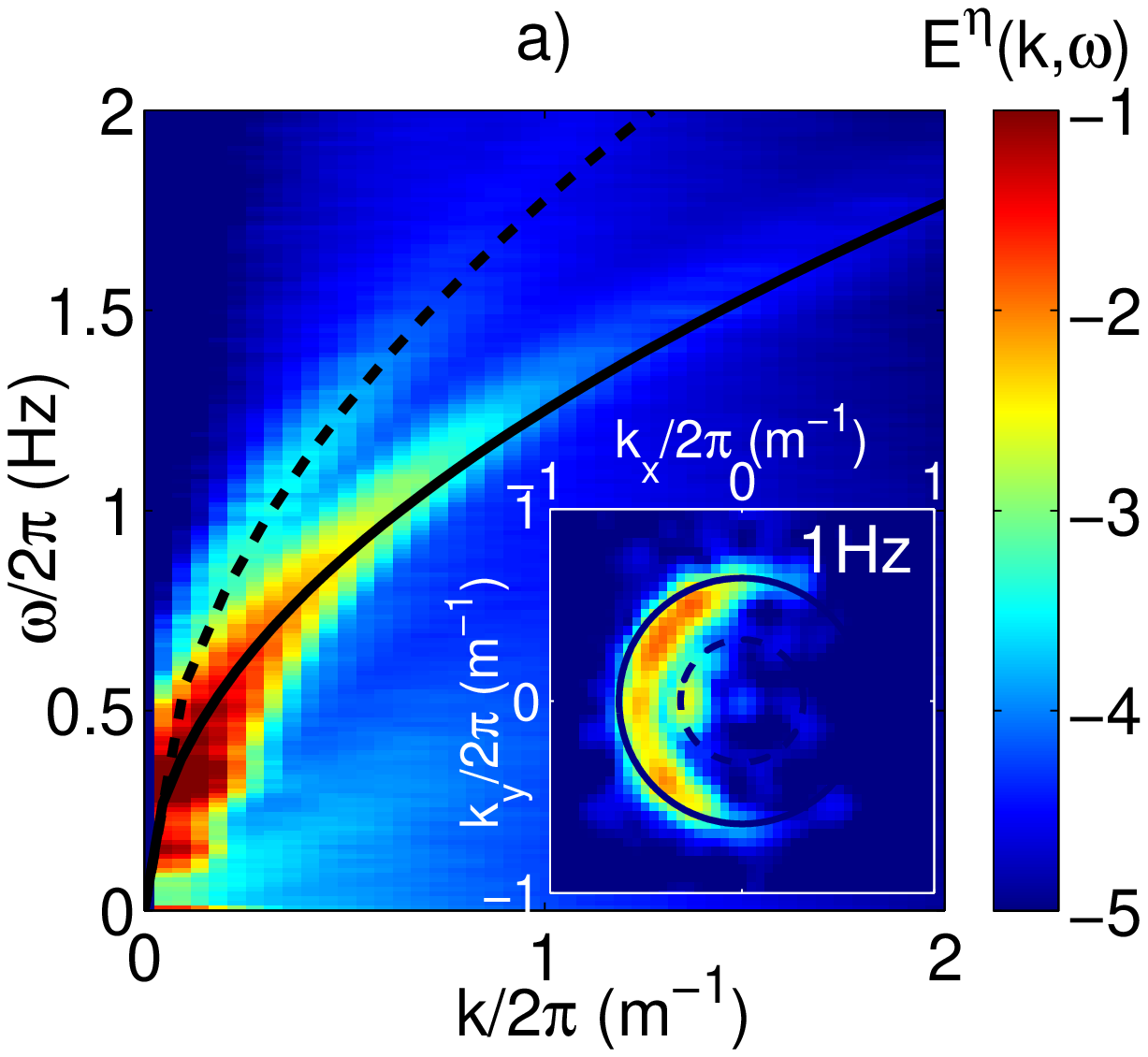}
\includegraphics[width=8cm]{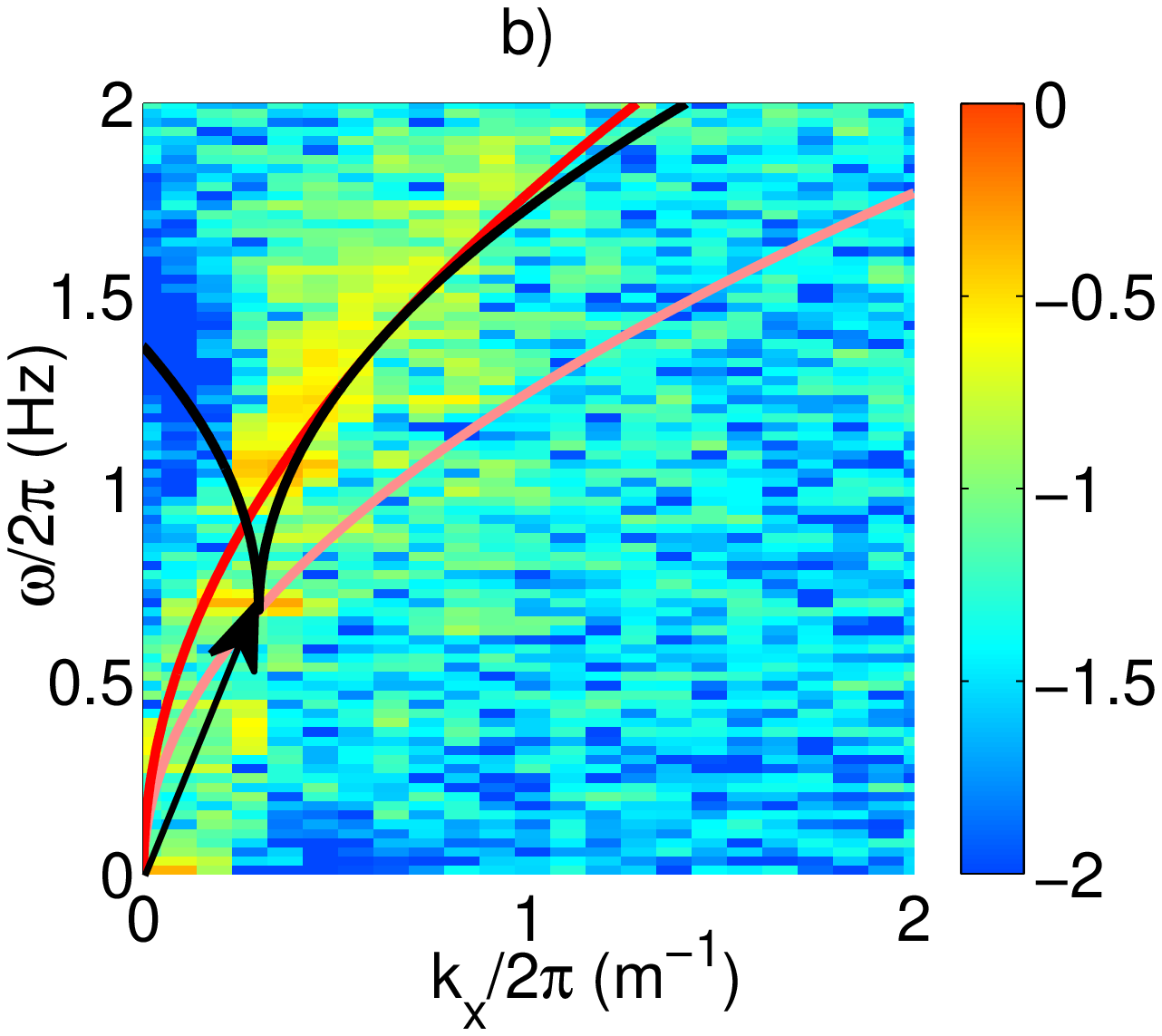}
\caption{(a) Power spectrum $E^\eta(k,\omega)$ of surface gravity waves measured by Leckler {\it et al.} \cite{Leckler2015} in the Black Sea. Inset, cross-section of the complete spectrum at a given frequency of $1$~Hz. The solid line is the linear dispersion relation and the dashed line is the second harmonic. (b) Cross-section along $\mathbf{k_2},\omega_2$ ($k_y/2\pi=0 m^-1$) of the 3D bicoherence $B^{k,\omega}_3(\mathbf{k_2},\mathbf{k_3},\omega_2,\omega_3)$ for a given wave $(\mathbf{k_2},\omega_2)$ computed from the same dataset. The red and black curves are similar to ones shown in Fig.~\ref{fig6}. Exact solutions are in the region where the black curve intersects the red one. A high correlation intensity is observed at this region, suggesting the occurrence of quasi-resonant interaction between linear waves and harmonic.}
\label{fig9}
\end{figure}

Thanks to the dataset of in-situ stereoscopic measurements in the Black Sea by Leckler {\it et al.}\cite{Leckler2015}, we are able to reproduce and compare the previous correlation analysis with actual sea waves. Surface gravity waves have been measured using a stereoscopic measurement based on a direct cross-correlation of the sea using natural patterns generated by the surface roughness (ripples) and impurities under diffuse lightning. The measurement is performed over a field of about $15\times 20$~m$^2$ during 30 minutes. This larger size allow by Nature is thus far enough to diminish strongly the effects of surface pollution caused by Marangoni waves as discussed above. Figure~\ref{fig9} a) shows a reproduction of their measured power spectrum $E^\eta(\mathbf{k},\omega)$. Energy is present at frequencies ranging from $0.3$~Hz to roughly $2$~Hz. Like for the experiment presented above, the energy is mainly distributed around the linear dispersion relation but a small fraction of the energy is on the second harmonic. The main difference is that waves are quite anisotropic (see inset that shows the spatial distribution of the energy at $1$~Hz, showing that only negative $k_x$ are present) instead of an isotropic system as in our experiment. 

Figure~\ref{fig9}(b) displays a cross-section of the 3D bicoherence similarly as Fig.~\ref{fig8}(a) above at $k_y/2\pi=0$~m$^{-1}$. Although the estimator is less converged than for the experiment due to a smaller amount of data, we clearly observe the presence of a high correlation area in the proximity of the black and the red curve (exact solutions). 
Like for our experiment, no quasi-resonant interactions between the two linear branches (pink and black curves) can be seen. Finally, the main difference with the experiment is the absence of the unexplained correlations on the negative parts of $(\mathbf{k_2},\omega_2)$. This absence supports the suggestion of that they may be related to standing waves in our experiment since these are not present in this field dataset.

\section{4-wave interactions}
\label{4w}
\begin{figure}[!htb]
\includegraphics[width=8cm]{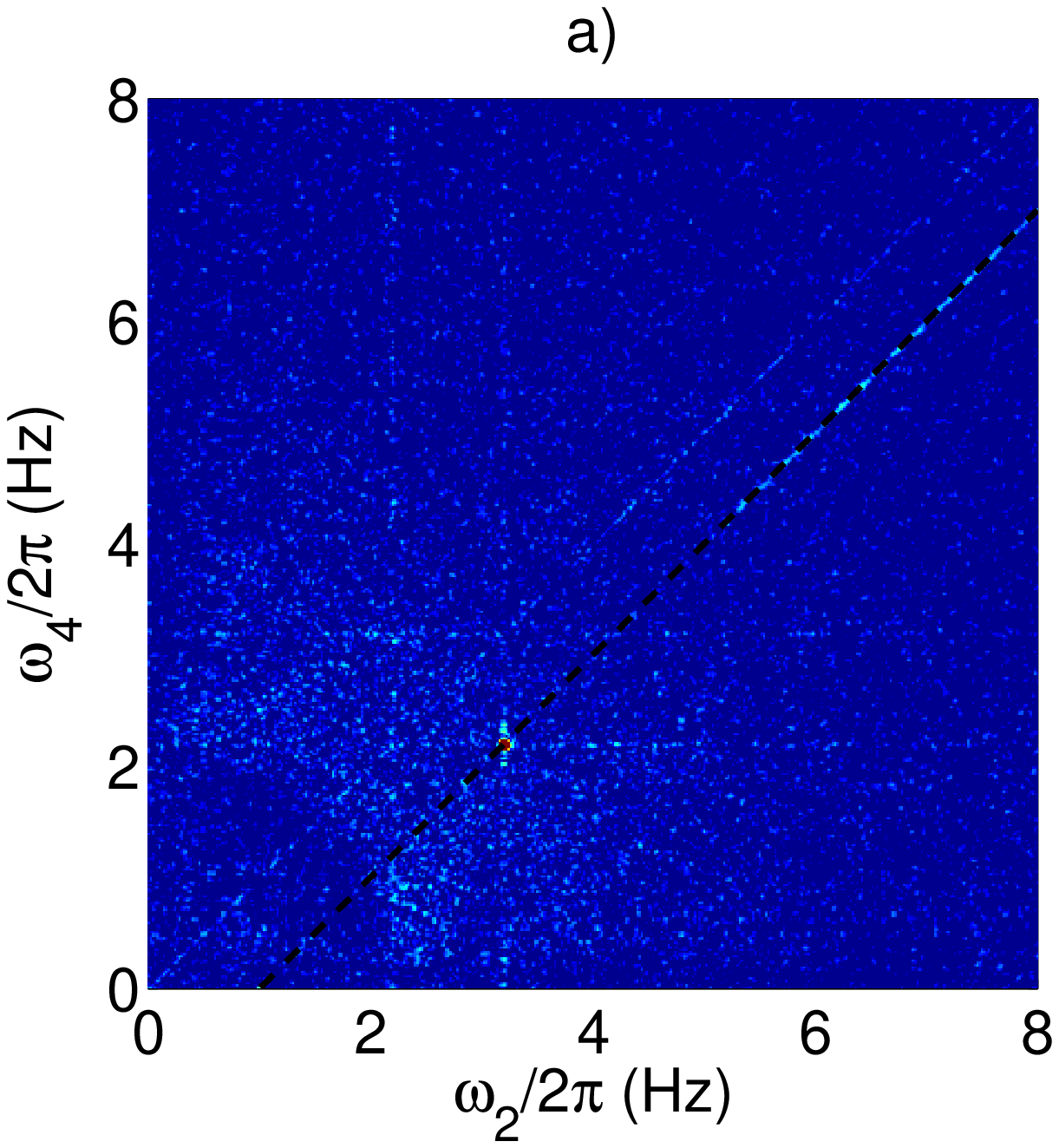}
\includegraphics[width=8cm]{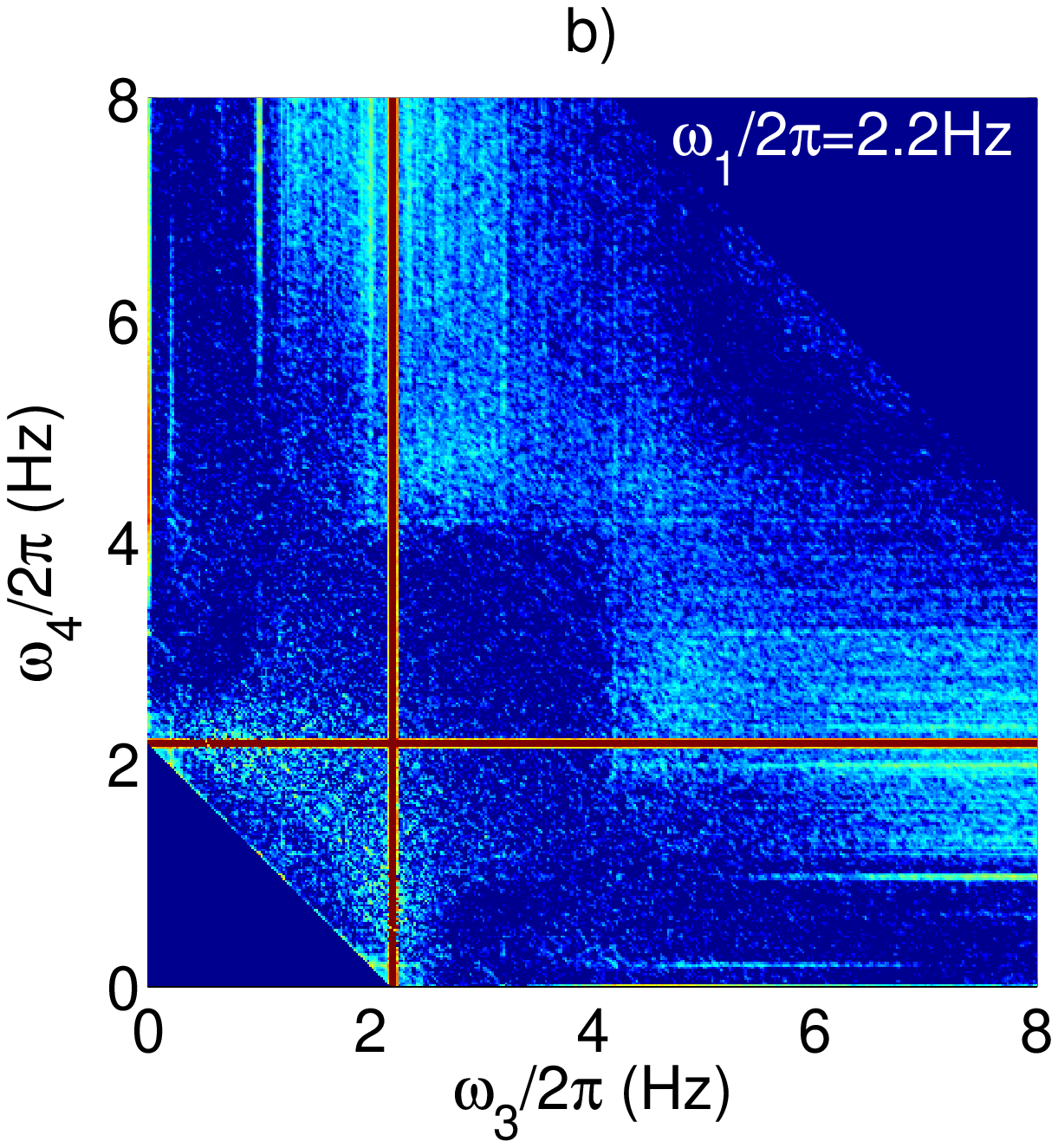}
\includegraphics[width=8cm]{cbar115.eps}
\caption{(a) Normalized fourth order correlations $C_4^{\omega}(\omega_1,\omega_2,\omega_3,\omega_4)$ for two given waves $\omega_1/2\pi=2.2$~Hz and $\omega_3/2\pi=3.2$~Hz in the experiment. 
A very weak value of correlations is present on the resonant line $\omega_1+\omega_2=\omega_3+\omega_4$ (dashed line) at the highest frequencies ($\omega_2/2\pi>5$~Hz). At low frequencies where linear waves exists, there is no evidence of 4-wave interactions at the level of convergence which is about $10^{-2.5}$. 
(b) Tricoherence $B_4^{\omega}(\omega_1,\omega_3,\omega_4)$. Like in (a), only a weak correlation is observed at the highest frequencies. The light blue on the bottom left corner of the figure ($\omega_3/2\pi<2.2$~Hz) is not statically converged. The two red lines correspond to the trivial solution $\omega_3=\omega_1$ and $\omega_4=\omega_2$  (or $\omega_4=\omega_1$ and  $\omega_3=\omega_2$).}
\label{fig10}
\end{figure}

The previous analysis shows that $3$-wave coupling exists between linear and harmonic waves, both in the experiment and in the sea. We now investigate of $4$-waves interactions that actually predicted to operate the turbulent cascade by the WTT. We compute the fourth order correlation in frequency normalized by the power spectrum. 
\begin{equation}
C^4_{\omega}(\omega_1,\omega_2,\omega_3,\omega_4)=\frac{\left\langle w^*(\omega_1)w^*(\omega_2)w(\omega_3)w(\omega_4)\right\rangle}{\sqrt{\left\langle \left|w(\omega_1)w(\omega_2)\right|^2\right\rangle\left\langle \left|w(\omega_3)w(\omega_4)\right|^2\right\rangle}}
\label{C4w}
\end{equation}
Figure \ref{fig10} a) shows this estimator for two arbitrary given frequencies $\omega_1/2\pi=2.2$~Hz and $\omega_3/2\pi=3.2$~Hz at which a significant value of the spectrum is observed. Unlike the third order correlation, their is no clear evidence of signal on the resonant line for this pair $(\omega_1,\omega_3)$. A significant correlation area is emerging but only at the highest frequencies ($\omega_2/2\pi\geq 5$~Hz). Its origin is unclear and may be due to noise as the spectrum is very weak at these frequencies and mostly at the noise level. 

In order to test more resonant configurations beyond this given pair, we compute the tricoherence 
\begin{equation}
T^4_{\omega}(\omega_1,\omega_3,\omega_4)=\frac{C^4_{\omega}(\omega_1,\omega_3+\omega_4-\omega_1,\omega_3,\omega_4)}{\sqrt{\left\langle \left|w(\omega_1)w(\omega_3+\omega_4-\omega_1)\right|^2\right\rangle\left\langle \left|w(\omega_3)w(\omega_4)\right|^2\right\rangle}}
\label{T4w}
\end{equation}
We give one value of the frequency $\omega_2$ to obtain 2D pictures. Figure \ref{fig10}(b) displays the tricoherence for $\omega_1/2\pi=2.2$~Hz. The two red lines are the trivial solutions corresponding to $\omega_3=\omega_1$ and $\omega_4=\omega_2$  (or $\omega_4=\omega_1$ and  $\omega_3=\omega_2$) and thus their value is not relevant. At low frequencies, where linear waves exists (with a significant level of energy in the spectrum), there is still no evidence of $4$-waves interaction since the observed coherence values are very weak. They are most likely not converged and they remain close to the statistical background noise. This pattern has been check for other given values of the frequency and leads to the conclusion that no $4$-waves correlations are observed at our convergence level for this experiment. 

\begin{figure}[!htb]
\includegraphics[width=8cm]{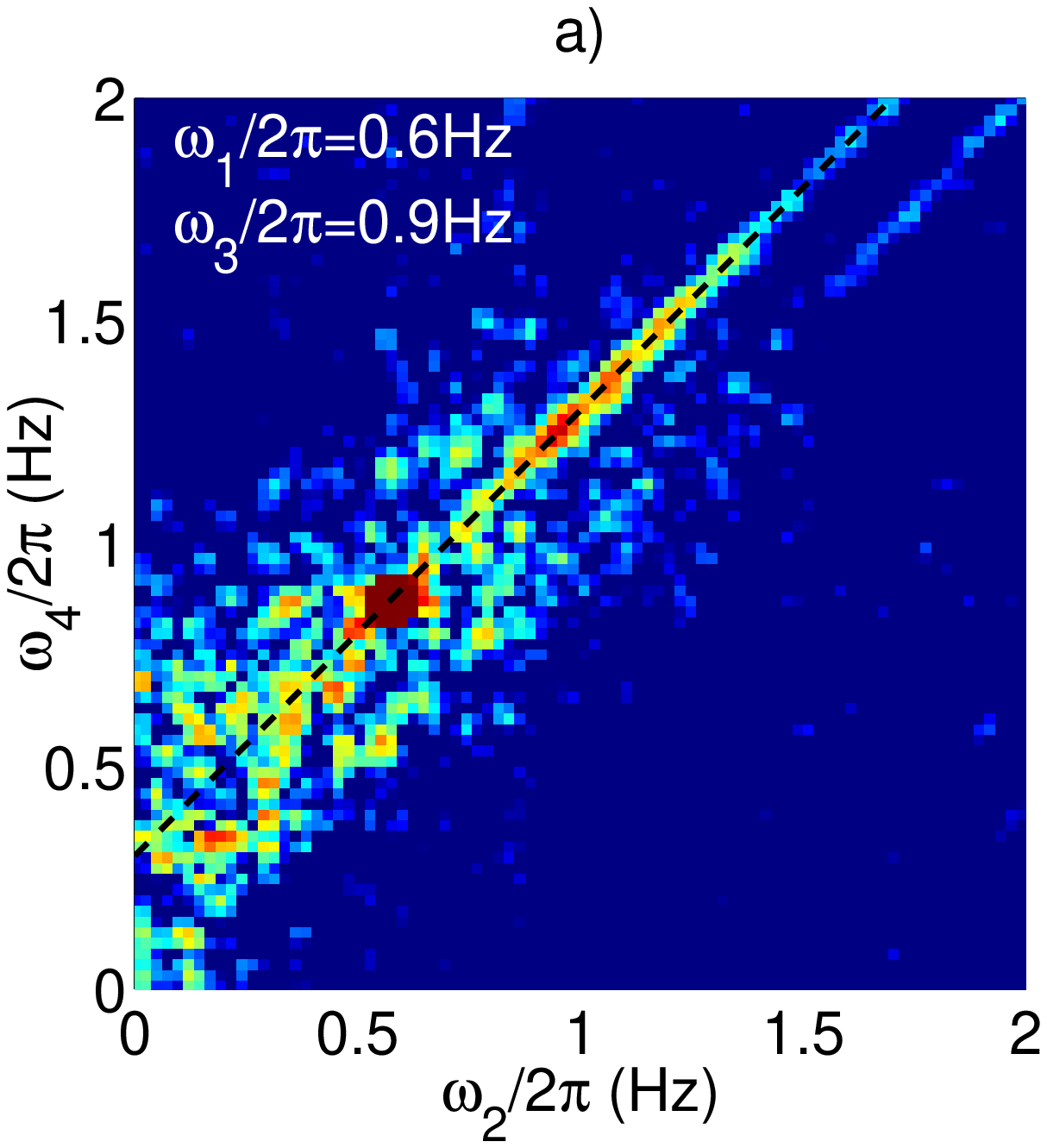}
\includegraphics[width=8cm]{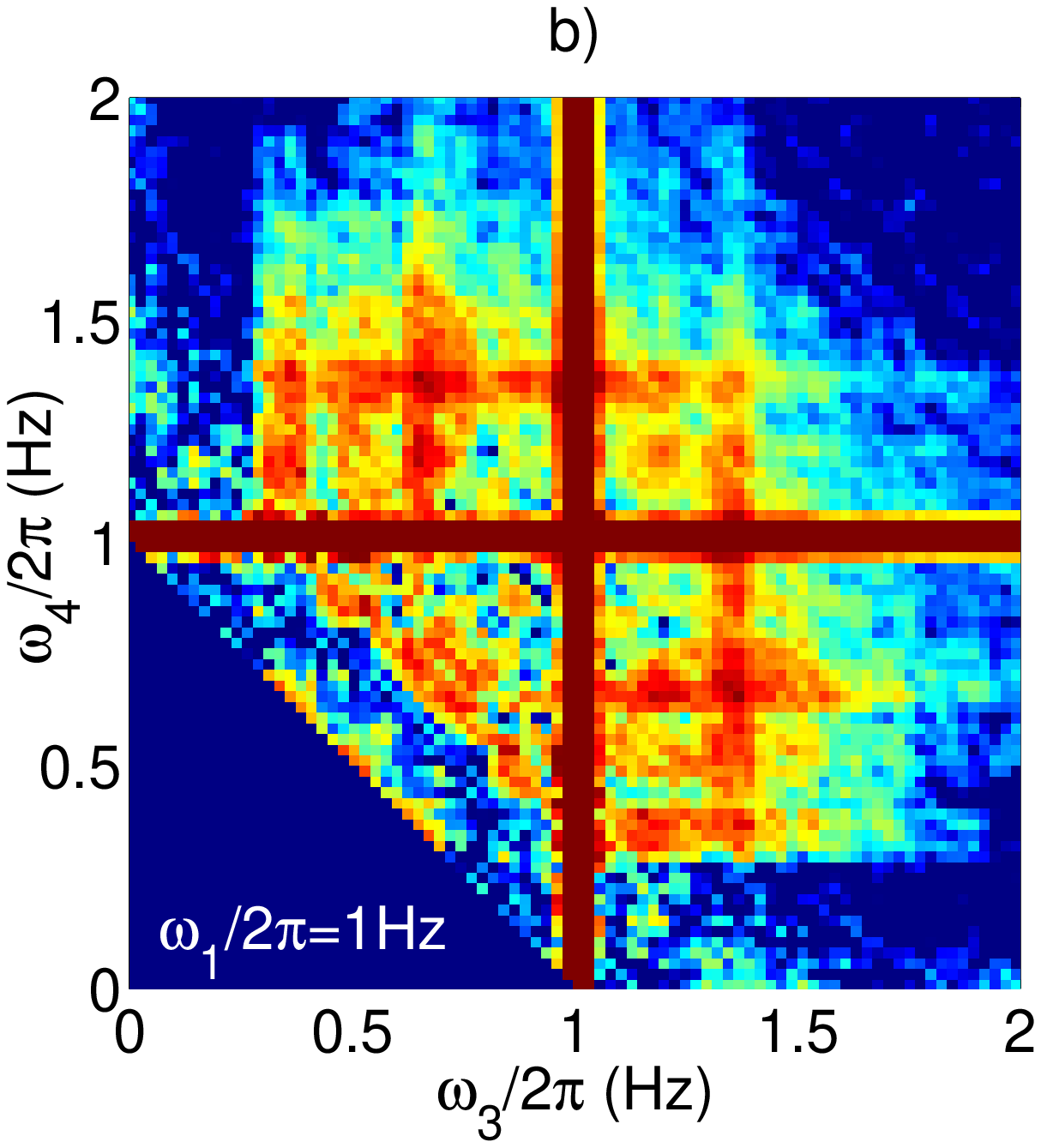}
\includegraphics[width=8cm]{cbar115.eps}
\caption{ Normalized fourth order correlation $C_4^{\omega}(\omega_1,\omega_2,\omega_3,\omega_4)$ computed for the in-situ dataset. Two arbitrary frequencies are given $\omega_1/2\pi=2.2$~Hz and $\omega_3/2\pi=3.2$~Hz. A significant level of statistically converged correlation is present along the resonant line $\omega_1+\omega_2=\omega_3+\omega_4$ (dashed line). b) Tricoherence $T_4^{\omega}(\omega_1,\omega_3,\omega_4)$ for $\omega_1/2\pi=1$~Hz. The two dark red lines correspond to the trivial solution $\omega_3=\omega_1$ and $\omega_4=\omega_2$  (or $\omega_4=\omega_1$ and  $\omega_3=\omega_2$). A significant level of correlation can be seen out of these trivial lines.}
\label{fig11}
\end{figure}

We then repeat the same analysis for the Black Sea dataset. Figure~\ref{fig11} shows a sample of both the fourth oder correlation $C^4_{\omega}$ and the tricoherence $T^4_{\omega}$. Although the statistical convergence is not as good as in the experiment a line of significant correlation levels shows up in Fig.~\ref{fig11}(a), in contrast with what was observed in the experiment. The tricoherence (fig\ref{fig11} b)) displays a significant value in a large part of the frequency space. In particular there is no special accumulation close to the trivial point $\omega_1=\omega_2=\omega_3=\omega_4$ (intersection of the two red lines) which mean that the wave coupling is not necessarily strongly local. We note some organization with lines, suggesting that a few modes are more effective (at $0.7$~Hz or $1.4$~Hz for instance). Due to the superposition of solutions in the frequency space, it is not possible to get more information from these plots. The next step is thus to do a space and time analysis like for the $3$-wave analysis. Unfortunately the number of realizations that permits to converge a statistical estimator decreases strongly in a full $(\mathbf k,\omega)$ domain due to the impossibility to perform spatial averaging. Thus, it does not permit to converge significantly the space and time tricoherence for which the correlation level is weaker than the one observed in the $3$-wave analysis.

\section{Discussion and concluding remarks} 
The correlation analysis that we presented confirms the presence of resonant interactions among pure gravity surface waves. However, the first results do not correspond to the theoretical expectations of $4$-wave interactions. Although no $3$-wave interactions are observed between linear waves as predicted by the theory, we actually observe $3$-wave coupling between the linear component and harmonic. These kinds of coupling are usually known as bound waves where a wave interacts with itself to form its own harmonic. However the resonances are not restricted to strict harmonics and we observe also quasi-resonant interactions that appear to be very active as well. These new solutions enhance the potential efficiency of these interactions among all possible bound waves. The remaining question is the contribution of the $3$-waves coupling in the generation of the global energy cascade. Indeed, despite strong correlations, the energy of the harmonic branch remains low (maximum of $10\%$). A bound wave resulting from the quadratic interaction of free waves, 3-wave interaction involving a bound wave can appear as a 4-free-wave interaction. This is actually the meaning of the normal-form reduction which is performed in the application of the weak turbulence theory~\cite{R9,Krasit}. The observed resonant coupling involving the second harmonics and near resonant bound waves maybe a privileged route for the global 4-wave process.

%

\begin{figure}[!htb]
\includegraphics[height=7cm]{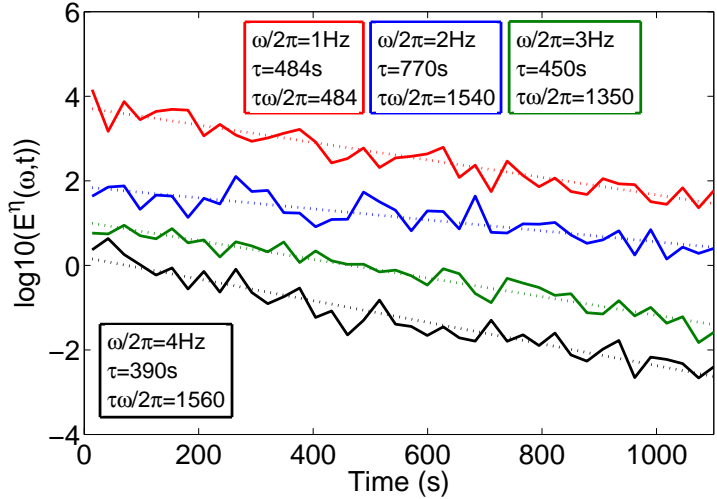}
\caption{Time evolution of the power spectrum $E^{\eta}(\omega,t)$ for four frequencies $1,2,3$and$4$~Hz. An exponential decrease is observed (dashed lines) and leads to a typical time scale $\tau$ following $e^{-\frac{t}{\tau}}$. We assume that the dissipative timescale $T^d$ is $T^{d}\approx \tau$.} 
\label{fig12}
\end{figure}

Surprisingly, the $4$th order correlation analysis in our experiment does not show traces of $4$-waves interactions that are supposedly at the core of the energy transfers for surface gravity waves. A turbulent-like spectrum is observed that suggests that an energy cascade is operating nonetheless. Two hypotheses may be proposed. The first one is that our convergence level remains too low to permit the observation of a possible very weak level of $4$-waves correlations. The second hypothesis is that the dissipation is too strong and breaks the scale separation that is the core of the theory. The typical $4-$wave non-linear time scale of energy transfers for this experiment, can be roughly estimated as $T^{NL}_{4w}=\frac{1}{\epsilon^{2(4-2)}}\frac{2\pi}{\omega}\approx 7000\frac{2\pi}{\omega}$ \cite{R2}. It may become larger than the typical time scale for the dissipation $T^{d}$. The latter decreases strongly with the presence of Marangoni waves and it has been estimated by studying the decrease of the energy when the forcing is stopped. Figure \ref{fig12} shows this evolution for four given frequencies.
We observe an exponential decrease $e^{-t\tau}$ that permits to define the dissipation time as $T^{d}\approx \tau$. We found an order of magnitude $T^{d}\approx 1000\frac{2\pi}{\omega}$ for the range of our measured waves. The prediction for the $3$-waves nonlinear time scale gives $T^{NL}_{3w}=\frac{1}{\epsilon^{2(3-2)}}\frac{2\pi}{\omega}\approx 80\frac{2\pi}{\omega}$. Thus we have  
\begin{equation}
T^{L} < T^{NL}_{3w} <  T^{d} < T^{NL}_{4w}
\end{equation}
and this supports the situation where only $3$-waves interactions are observed. From this observation, we may propose that the additional dissipation does not permit the usual waves resonant interactions and the cascade may thus be generated through $3$-waves interactions using the harmonic branch.

Since the scales are much larger in the ocean, the field measurements are less affected by the additional dissipation by the Marangoni waves. We then expect to observe classical wave turbulence through only $4$-waves interaction for weak enough waves if $T^{NL}_{4w}<T^d$. Indeed, the $4$-waves correlations show a signature of $4$-waves interactions for the Black Sea dataset. The mean level of these interaction remain lower (about $5\%$) than the $3$-waves interactions with the harmonic branch (about $25\%$) that are also observed in this regime. However, due to the weak energy level of the harmonic branch, it is possible that a significant part of the transfer occurs through these $4$-waves interactions. This observation is in agreement with the exponent of the power spectrum which is close to the theoretical prediction while keeping a very weak non-linearity with $\epsilon_p=0.02$ as seen in Fig.~\ref{fig4}. Nevertheless it should be noted that although the wave steepness is much lower in the Black Sea than in the experiment (by roughly one order of magnitude) the harmonic branch remains very visible in the spectrum and the 3-wave correlations level remains quite high. This observation suggests that, for a finite level of nonlinearity, the 3-wave resonant coupling with the harmonic that we describe here may actually be an important mechanism for the energy redistribution even when 4-wave coupling is operating.

To sum up, $4$-waves interactions seem to exist but their very low occurrence probability and their associated long time scale make them very sensitive to the presence of dissipation in the inertial range. This is especially the case of very weakly non-linear regimes where the non-linear interaction time become very high. Our observations suggest an additional, faster and thus more efficient 3-wave coupling mechanism for the energy transfers. The question is whether these 3-wave interactions can play a role in the turbulence cascade and replace the weak turbulence $4$-waves process. 



\begin{acknowledgments}
This project has received funding from the European Research Council (ERC) under the European Union's Horizon 2020 research and innovation programme (grant agreement No 647018-WATU). A.C. and the project were also supported by Fondation Simone et Cino Del Duca and Institut de France.\end{acknowledgments}

\bibliography{bibliogravity}

\begin{thebibliography}{34}%
\makeatletter
\providecommand \@ifxundefined [1]{%
 \@ifx{#1\undefined}
}%
\providecommand \@ifnum [1]{%
 \ifnum #1\expandafter \@firstoftwo
 \else \expandafter \@secondoftwo
 \fi
}%
\providecommand \@ifx [1]{%
 \ifx #1\expandafter \@firstoftwo
 \else \expandafter \@secondoftwo
 \fi
}%
\providecommand \natexlab [1]{#1}%
\providecommand \enquote  [1]{``#1''}%
\providecommand \bibnamefont  [1]{#1}%
\providecommand \bibfnamefont [1]{#1}%
\providecommand \citenamefont [1]{#1}%
\providecommand \href@noop [0]{\@secondoftwo}%
\providecommand \href [0]{\begingroup \@sanitize@url \@href}%
\providecommand \@href[1]{\@@startlink{#1}\@@href}%
\providecommand \@@href[1]{\endgroup#1\@@endlink}%
\providecommand \@sanitize@url [0]{\catcode `\\12\catcode `\$12\catcode
  `\&12\catcode `\#12\catcode `\^12\catcode `\_12\catcode `\%12\relax}%
\providecommand \@@startlink[1]{}%
\providecommand \@@endlink[0]{}%
\providecommand \url  [0]{\begingroup\@sanitize@url \@url }%
\providecommand \@url [1]{\endgroup\@href {#1}{\urlprefix }}%
\providecommand \urlprefix  [0]{URL }%
\providecommand \Eprint [0]{\href }%
\providecommand \doibase [0]{http://dx.doi.org/}%
\providecommand \selectlanguage [0]{\@gobble}%
\providecommand \bibinfo  [0]{\@secondoftwo}%
\providecommand \bibfield  [0]{\@secondoftwo}%
\providecommand \translation [1]{[#1]}%
\providecommand \BibitemOpen [0]{}%
\providecommand \bibitemStop [0]{}%
\providecommand \bibitemNoStop [0]{.\EOS\space}%
\providecommand \EOS [0]{\spacefactor3000\relax}%
\providecommand \BibitemShut  [1]{\csname bibitem#1\endcsname}%
\let\auto@bib@innerbib\@empty
\bibitem [{\citenamefont {Zakharov}\ \emph {et~al.}(1992)\citenamefont
  {Zakharov}, \citenamefont {L'vov},\ and\ \citenamefont {Falkovich}}]{R1}%
  \BibitemOpen
  \bibfield  {author} {\bibinfo {author} {\bibfnamefont {V.~E.}\ \bibnamefont
  {Zakharov}}, \bibinfo {author} {\bibfnamefont {V.~S.}\ \bibnamefont {L'vov}},
  \ and\ \bibinfo {author} {\bibfnamefont {G.}~\bibnamefont {Falkovich}},\
  }\href@noop {} {\emph {\bibinfo {title} {Kolmogorov Spectra of Turbulence}}}\
  (\bibinfo  {publisher} {Springer},\ \bibinfo {address} {Berlin},\ \bibinfo
  {year} {1992})\BibitemShut {NoStop}%
\bibitem [{\citenamefont {Nazarenko}(2011)}]{R2}%
  \BibitemOpen
  \bibfield  {author} {\bibinfo {author} {\bibfnamefont {S.}~\bibnamefont
  {Nazarenko}},\ }\href@noop {} {\emph {\bibinfo {title} {Wave Turbulence}}}\
  (\bibinfo  {publisher} {Springer},\ \bibinfo {address} {Berlin},\ \bibinfo
  {year} {2011})\BibitemShut {NoStop}%
\bibitem [{\citenamefont {Newell}\ and\ \citenamefont {Rumpf}(2011)}]{R3}%
  \BibitemOpen
  \bibfield  {author} {\bibinfo {author} {\bibfnamefont {A.~C.}\ \bibnamefont
  {Newell}}\ and\ \bibinfo {author} {\bibfnamefont {B.}~\bibnamefont {Rumpf}},\
  }\bibfield  {title} {\enquote {\bibinfo {title} {Wave turbulence},}\
  }\href@noop {} {\bibfield  {journal} {\bibinfo  {journal} {Ann. Rev. Fluid
  Mech.}\ }\textbf {\bibinfo {volume} {43}} (\bibinfo {year}
  {2011})}\BibitemShut {NoStop}%
\bibitem [{\citenamefont {Zakharov}(1984)}]{zakharov1984kolmogorov}%
  \BibitemOpen
  \bibfield  {author} {\bibinfo {author} {\bibfnamefont {VE}~\bibnamefont
  {Zakharov}},\ }\bibfield  {title} {\enquote {\bibinfo {title} {Kolmogorov
  spectra in weak turbulence problems},}\ }in\ \href@noop {} {\emph {\bibinfo
  {booktitle} {Basic Plasma Physics: Selected Chapters, Handbook of Plasma
  Physics, Volume 1}}}\ (\bibinfo {year} {1984})\ p.~\bibinfo {pages}
  {3}\BibitemShut {NoStop}%
\bibitem [{\citenamefont {Hasselmann}(1962)}]{R9}%
  \BibitemOpen
  \bibfield  {author} {\bibinfo {author} {\bibfnamefont {K.}~\bibnamefont
  {Hasselmann}},\ }\bibfield  {title} {\enquote {\bibinfo {title} {On the
  non-linear energy transfer in gravity-wave spectrum. part 1. general
  theory},}\ }\href@noop {} {\bibfield  {journal} {\bibinfo  {journal} {J.
  Fluid Mech.}\ }\textbf {\bibinfo {volume} {12}},\ \bibinfo {pages} {481--500}
  (\bibinfo {year} {1962})}\BibitemShut {NoStop}%
\bibitem [{\citenamefont {Berhanu}\ and\ \citenamefont
  {Falcon}(2013)}]{Berhanu2013}%
  \BibitemOpen
  \bibfield  {author} {\bibinfo {author} {\bibfnamefont {Michael}\ \bibnamefont
  {Berhanu}}\ and\ \bibinfo {author} {\bibfnamefont {Eric}\ \bibnamefont
  {Falcon}},\ }\bibfield  {title} {\enquote {\bibinfo {title}
  {Space-time-resolved capillary wave turbulence},}\ }\href {\doibase
  10.1103/PhysRevE.87.033003} {\bibfield  {journal} {\bibinfo  {journal} {Phys.
  Rev. E}\ }\textbf {\bibinfo {volume} {87}},\ \bibinfo {pages} {033003}
  (\bibinfo {year} {2013})}\BibitemShut {NoStop}%
\bibitem [{\citenamefont {Falcon}\ \emph {et~al.}(2007)\citenamefont {Falcon},
  \citenamefont {Laroche},\ and\ \citenamefont {Fauve}}]{Falcon2007}%
  \BibitemOpen
  \bibfield  {author} {\bibinfo {author} {\bibfnamefont {Eric}\ \bibnamefont
  {Falcon}}, \bibinfo {author} {\bibfnamefont {Claude}\ \bibnamefont
  {Laroche}}, \ and\ \bibinfo {author} {\bibfnamefont {S.}~\bibnamefont
  {Fauve}},\ }\bibfield  {title} {\enquote {\bibinfo {title} {Observation of
  gravity-capillary wave turbulence},}\ }\href {\doibase
  10.1103/PhysRevLett.98.094503} {\bibfield  {journal} {\bibinfo  {journal}
  {Phys. Rev. Lett.}\ }\textbf {\bibinfo {volume} {98}},\ \bibinfo {pages}
  {094503} (\bibinfo {year} {2007})}\BibitemShut {NoStop}%
\bibitem [{\citenamefont {Falcon}\ \emph {et~al.}(2009)\citenamefont {Falcon},
  \citenamefont {Bortolozzo},\ and\ \citenamefont {Fauve}}]{Falcon2009}%
  \BibitemOpen
  \bibfield  {author} {\bibinfo {author} {\bibfnamefont {Eric}\ \bibnamefont
  {Falcon}}, \bibinfo {author} {\bibfnamefont {U.}~\bibnamefont {Bortolozzo}},
  \ and\ \bibinfo {author} {\bibfnamefont {S.}~\bibnamefont {Fauve}},\
  }\bibfield  {title} {\enquote {\bibinfo {title} {Capillary wave turbulence on
  a spherical fluid surface in low gravity},}\ }\href {\doibase
  10.1209/0295-5075/86/14002} {\bibfield  {journal} {\bibinfo  {journal} {EPL
  (Europhysics Letters)}\ }\textbf {\bibinfo {volume} {86}},\ \bibinfo {pages}
  {14002} (\bibinfo {year} {2009})}\BibitemShut {NoStop}%
\bibitem [{\citenamefont {Deike}\ \emph {et~al.}(2014)\citenamefont {Deike},
  \citenamefont {Fuster}, \citenamefont {Berhanu},\ and\ \citenamefont
  {Falcon}}]{Deike2014}%
  \BibitemOpen
  \bibfield  {author} {\bibinfo {author} {\bibfnamefont {Luc}\ \bibnamefont
  {Deike}}, \bibinfo {author} {\bibfnamefont {Daniel}\ \bibnamefont {Fuster}},
  \bibinfo {author} {\bibfnamefont {Michael}\ \bibnamefont {Berhanu}}, \ and\
  \bibinfo {author} {\bibfnamefont {Eric}\ \bibnamefont {Falcon}},\ }\bibfield
  {title} {\enquote {\bibinfo {title} {Direct numerical simulations of
  capillary wave turbulence},}\ }\href {\doibase
  10.1103/PhysRevLett.112.234501} {\bibfield  {journal} {\bibinfo  {journal}
  {Phys. Rev. Lett.}\ }\textbf {\bibinfo {volume} {112}},\ \bibinfo {pages}
  {234501} (\bibinfo {year} {2014})},\ \Eprint {http://arxiv.org/abs/1406.0795}
  {1406.0795} \BibitemShut {NoStop}%
\bibitem [{\citenamefont {Cobelli}\ \emph {et~al.}(2011)\citenamefont
  {Cobelli}, \citenamefont {Przadka}, \citenamefont {Petitjeans}, \citenamefont
  {Lagubeau}, \citenamefont {Pagneux},\ and\ \citenamefont {Maurel}}]{CoPr11}%
  \BibitemOpen
  \bibfield  {author} {\bibinfo {author} {\bibfnamefont {P.}~\bibnamefont
  {Cobelli}}, \bibinfo {author} {\bibfnamefont {A.}~\bibnamefont {Przadka}},
  \bibinfo {author} {\bibfnamefont {P.}~\bibnamefont {Petitjeans}}, \bibinfo
  {author} {\bibfnamefont {G.}~\bibnamefont {Lagubeau}}, \bibinfo {author}
  {\bibfnamefont {V.}~\bibnamefont {Pagneux}}, \ and\ \bibinfo {author}
  {\bibfnamefont {A.}~\bibnamefont {Maurel}},\ }\bibfield  {title} {\enquote
  {\bibinfo {title} {Different regimes for water wave turbulence},}\
  }\href@noop {} {\bibfield  {journal} {\bibinfo  {journal} {Phys. Rev. Lett.}\
  }\textbf {\bibinfo {volume} {107}},\ \bibinfo {pages} {214503} (\bibinfo
  {year} {2011})}\BibitemShut {NoStop}%
\bibitem [{\citenamefont {Nazarenko}\ \emph {et~al.}(2009)\citenamefont
  {Nazarenko}, \citenamefont {Lukaschuk}, \citenamefont {McLelland},\ and\
  \citenamefont {Denissenko}}]{Nazarenko2009}%
  \BibitemOpen
  \bibfield  {author} {\bibinfo {author} {\bibfnamefont {Sergey}\ \bibnamefont
  {Nazarenko}}, \bibinfo {author} {\bibfnamefont {S.}~\bibnamefont
  {Lukaschuk}}, \bibinfo {author} {\bibfnamefont {Stuart}\ \bibnamefont
  {McLelland}}, \ and\ \bibinfo {author} {\bibfnamefont {Petr}\ \bibnamefont
  {Denissenko}},\ }\bibfield  {title} {\enquote {\bibinfo {title} {Statistics
  of surface gravity wave turbulence in the space and time domains},}\ }\href
  {\doibase 10.1017/S0022112009991820} {\bibfield  {journal} {\bibinfo
  {journal} {J. Fluid Mech.}\ }\textbf {\bibinfo {volume} {642}},\ \bibinfo
  {pages} {395} (\bibinfo {year} {2009})}\BibitemShut {NoStop}%
\bibitem [{\citenamefont {Deike}\ \emph {et~al.}(2015)\citenamefont {Deike},
  \citenamefont {Miquel}, \citenamefont {Jamin}, \citenamefont {Semin},
  \citenamefont {Berhanu}, \citenamefont {Falcon},\ and\ \citenamefont
  {Bonnefoy}}]{Deike2015}%
  \BibitemOpen
  \bibfield  {author} {\bibinfo {author} {\bibfnamefont {Luc}\ \bibnamefont
  {Deike}}, \bibinfo {author} {\bibfnamefont {B}~\bibnamefont {Miquel}},
  \bibinfo {author} {\bibfnamefont {T}~\bibnamefont {Jamin}}, \bibinfo {author}
  {\bibfnamefont {B}~\bibnamefont {Semin}}, \bibinfo {author} {\bibfnamefont
  {Michael}\ \bibnamefont {Berhanu}}, \bibinfo {author} {\bibfnamefont {Eric}\
  \bibnamefont {Falcon}}, \ and\ \bibinfo {author} {\bibfnamefont
  {F}~\bibnamefont {Bonnefoy}},\ }\bibfield  {title} {\enquote {\bibinfo
  {title} {Role of the basin boundary conditions in gravity wave turbulence},}\
  }\href {\doibase 10.1017/jfm.2015.494} {\bibfield  {journal} {\bibinfo
  {journal} {J. Fluid Mech.}\ }\textbf {\bibinfo {volume} {781}},\ \bibinfo
  {pages} {196----225} (\bibinfo {year} {2015})}\BibitemShut {NoStop}%
\bibitem [{\citenamefont {Denissenko}\ \emph {et~al.}(2007)\citenamefont
  {Denissenko}, \citenamefont {Lukaschuk},\ and\ \citenamefont
  {Nazarenko}}]{Denissenko2007}%
  \BibitemOpen
  \bibfield  {author} {\bibinfo {author} {\bibfnamefont {Petr}\ \bibnamefont
  {Denissenko}}, \bibinfo {author} {\bibfnamefont {S.}~\bibnamefont
  {Lukaschuk}}, \ and\ \bibinfo {author} {\bibfnamefont {Sergey}\ \bibnamefont
  {Nazarenko}},\ }\bibfield  {title} {\enquote {\bibinfo {title} {Gravity wave
  turbulence in a laboratory flume},}\ }\href {\doibase
  10.1103/PhysRevLett.99.014501} {\bibfield  {journal} {\bibinfo  {journal}
  {Phys. Rev. Lett.}\ }\textbf {\bibinfo {volume} {99}},\ \bibinfo {pages}
  {014501} (\bibinfo {year} {2007})}\BibitemShut {NoStop}%
\bibitem [{\citenamefont {Kartashova}(1998)}]{Kartashova1998}%
  \BibitemOpen
  \bibfield  {author} {\bibinfo {author} {\bibfnamefont {Elena}\ \bibnamefont
  {Kartashova}},\ }\bibfield  {title} {\enquote {\bibinfo {title} {Wave
  resonances in systems with discrete spectra},}\ }\href@noop {} {\bibfield
  {journal} {\bibinfo  {journal} {Am. Math. Soc. Trans.}\ ,\ \bibinfo {pages}
  {95--130}} (\bibinfo {year} {1998})}\BibitemShut {NoStop}%
\bibitem [{\citenamefont {Pushkarev}(1999)}]{Pushkarev1999}%
  \BibitemOpen
  \bibfield  {author} {\bibinfo {author} {\bibfnamefont {A}~\bibnamefont
  {Pushkarev}},\ }\bibfield  {title} {\enquote {\bibinfo {title} {On the
  kolmogorov and frozen turbulence in numerical simulation of capillary
  waves},}\ }\href {\doibase 10.1016/S0997-7546(99)80032-6} {\bibfield
  {journal} {\bibinfo  {journal} {Eur. J. Mech., B/Fluids}\ }\textbf {\bibinfo
  {volume} {18}},\ \bibinfo {pages} {345--351} (\bibinfo {year}
  {1999})}\BibitemShut {NoStop}%
\bibitem [{\citenamefont {Nazarenko}(2013)}]{Nazarenko2013}%
  \BibitemOpen
  \bibfield  {author} {\bibinfo {author} {\bibfnamefont {Sergey}\ \bibnamefont
  {Nazarenko}},\ }\bibfield  {title} {\enquote {\bibinfo {title} {Sandpile
  behaviour in discrete water-wave turbulence},}\ }\href {\doibase
  10.1088/1742-5468/2006/02/L02002} {\bibfield  {journal} {\bibinfo  {journal}
  {J. Stat. Mech.: Theory and Experiment}\ }\textbf {\bibinfo {volume}
  {02002}},\ \bibinfo {pages} {1--8} (\bibinfo {year} {2013})},\ \Eprint
  {http://arxiv.org/abs/0510054} {0510054 [nlin]} \BibitemShut {NoStop}%
\bibitem [{\citenamefont {Miquel}\ \emph {et~al.}(2014)\citenamefont {Miquel},
  \citenamefont {Alexakis},\ and\ \citenamefont {Mordant}}]{R23}%
  \BibitemOpen
  \bibfield  {author} {\bibinfo {author} {\bibfnamefont {B.}~\bibnamefont
  {Miquel}}, \bibinfo {author} {\bibfnamefont {A.}~\bibnamefont {Alexakis}}, \
  and\ \bibinfo {author} {\bibfnamefont {N.}~\bibnamefont {Mordant}},\
  }\bibfield  {title} {\enquote {\bibinfo {title} {Role of dissipation in
  flexural wave turbulence: from experimental spectrum to kolmogorov-zakharov
  spectrum},}\ }\href@noop {} {\bibfield  {journal} {\bibinfo  {journal} {Phys.
  Rev. E}\ }\textbf {\bibinfo {volume} {89}},\ \bibinfo {pages} {062925}
  (\bibinfo {year} {2014})}\BibitemShut {NoStop}%
\bibitem [{\citenamefont {Humbert}\ \emph {et~al.}(2013)\citenamefont
  {Humbert}, \citenamefont {Cadot}, \citenamefont {D\"uring}, \citenamefont
  {Josserand}, \citenamefont {Rica},\ and\ \citenamefont {Touz\'e}}]{Humbert}%
  \BibitemOpen
  \bibfield  {author} {\bibinfo {author} {\bibfnamefont {T.}~\bibnamefont
  {Humbert}}, \bibinfo {author} {\bibfnamefont {O.}~\bibnamefont {Cadot}},
  \bibinfo {author} {\bibfnamefont {G.}~\bibnamefont {D\"uring}}, \bibinfo
  {author} {\bibfnamefont {C.}~\bibnamefont {Josserand}}, \bibinfo {author}
  {\bibfnamefont {S.}~\bibnamefont {Rica}}, \ and\ \bibinfo {author}
  {\bibfnamefont {C.}~\bibnamefont {Touz\'e}},\ }\bibfield  {title} {\enquote
  {\bibinfo {title} {Wave turbulence in vibrating plates : the effect of
  damping},}\ }\href@noop {} {\bibfield  {journal} {\bibinfo  {journal} {EPL}\
  }\textbf {\bibinfo {volume} {102}},\ \bibinfo {pages} {30002} (\bibinfo
  {year} {2013})}\BibitemShut {NoStop}%
\bibitem [{\citenamefont {Alpers}\ and\ \citenamefont
  {H{\"{u}}hnerfuss}(1989)}]{Alpers1989}%
  \BibitemOpen
  \bibfield  {author} {\bibinfo {author} {\bibfnamefont {Werner}\ \bibnamefont
  {Alpers}}\ and\ \bibinfo {author} {\bibfnamefont {Heinrich}\ \bibnamefont
  {H{\"{u}}hnerfuss}},\ }\bibfield  {title} {\enquote {\bibinfo {title} {The
  damping of ocean waves by surface films: A new look at an old problem},}\
  }\href {\doibase 10.1029/JC094iC05p06251} {\bibfield  {journal} {\bibinfo
  {journal} {J. Geophys. Res.}\ }\textbf {\bibinfo {volume} {94}},\ \bibinfo
  {pages} {6251} (\bibinfo {year} {1989})}\BibitemShut {NoStop}%
\bibitem [{\citenamefont {Behroozi}\ \emph {et~al.}(2007)\citenamefont
  {Behroozi}, \citenamefont {Cordray},\ and\ \citenamefont
  {Griffin}}]{Behroozi2007}%
  \BibitemOpen
  \bibfield  {author} {\bibinfo {author} {\bibfnamefont {F}~\bibnamefont
  {Behroozi}}, \bibinfo {author} {\bibfnamefont {Kimberly}\ \bibnamefont
  {Cordray}}, \ and\ \bibinfo {author} {\bibfnamefont {William}\ \bibnamefont
  {Griffin}},\ }\bibfield  {title} {\enquote {\bibinfo {title} {The calming
  effect of oil on water},}\ }\href {\doibase 10.1119/1.2710482} {\bibfield
  {journal} {\bibinfo  {journal} {Am. J. Phys.}\ }\textbf {\bibinfo {volume}
  {75}},\ \bibinfo {pages} {407} (\bibinfo {year} {2007})}\BibitemShut
  {NoStop}%
\bibitem [{\citenamefont {Przadka}\ \emph {et~al.}(2011)\citenamefont
  {Przadka}, \citenamefont {Cabane}, \citenamefont {Pagneux}, \citenamefont
  {Maurel},\ and\ \citenamefont {Petitjeans}}]{Przadka2011}%
  \BibitemOpen
  \bibfield  {author} {\bibinfo {author} {\bibfnamefont {A.}~\bibnamefont
  {Przadka}}, \bibinfo {author} {\bibfnamefont {B.}~\bibnamefont {Cabane}},
  \bibinfo {author} {\bibfnamefont {V.}~\bibnamefont {Pagneux}}, \bibinfo
  {author} {\bibfnamefont {A.}~\bibnamefont {Maurel}}, \ and\ \bibinfo {author}
  {\bibfnamefont {P.}~\bibnamefont {Petitjeans}},\ }\bibfield  {title}
  {\enquote {\bibinfo {title} {Fourier transform profilometry for water waves:
  how to achieve clean water attenuation with diffusive reflection at the water
  surface?}}\ }\href {\doibase 10.1007/s00348-011-1240-x} {\bibfield  {journal}
  {\bibinfo  {journal} {Exp. Fluids}\ }\textbf {\bibinfo {volume} {52}},\
  \bibinfo {pages} {519--527} (\bibinfo {year} {2011})}\BibitemShut {NoStop}%
\bibitem [{\citenamefont {Aubourg}\ and\ \citenamefont
  {Mordant}(2015)}]{Aubourg2015}%
  \BibitemOpen
  \bibfield  {author} {\bibinfo {author} {\bibfnamefont {Quentin}\ \bibnamefont
  {Aubourg}}\ and\ \bibinfo {author} {\bibfnamefont {N.}~\bibnamefont
  {Mordant}},\ }\bibfield  {title} {\enquote {\bibinfo {title} {Nonlocal
  resonances in weak turbulence of gravity-capillary waves},}\ }\href {\doibase
  10.1103/PhysRevLett.114.144501} {\bibfield  {journal} {\bibinfo  {journal}
  {Phys. Rev. Lett.}\ }\textbf {\bibinfo {volume} {114}},\ \bibinfo {pages}
  {1--5} (\bibinfo {year} {2015})}\BibitemShut {NoStop}%
\bibitem [{\citenamefont {Aubourg}\ and\ \citenamefont
  {Mordant}(2016)}]{Aubourg2016}%
  \BibitemOpen
  \bibfield  {author} {\bibinfo {author} {\bibfnamefont {Quentin}\ \bibnamefont
  {Aubourg}}\ and\ \bibinfo {author} {\bibfnamefont {Nicolas}\ \bibnamefont
  {Mordant}},\ }\bibfield  {title} {\enquote {\bibinfo {title} {Investigation
  of resonances in gravity-capillary wave turbulence},}\ }\href {\doibase
  10.1103/PhysRevFluids.1.023701} {\bibfield  {journal} {\bibinfo  {journal}
  {Phys. Rev. Fluids}\ }\textbf {\bibinfo {volume} {1}},\ \bibinfo {pages}
  {023701} (\bibinfo {year} {2016})},\ \Eprint
  {http://arxiv.org/abs/1605.04091} {1605.04091} \BibitemShut {NoStop}%
\bibitem [{\citenamefont {Aubourg}(2016)}]{AubourgPhD}%
  \BibitemOpen
  \bibfield  {author} {\bibinfo {author} {\bibfnamefont {Quentin}\ \bibnamefont
  {Aubourg}},\ }\emph {\bibinfo {title} {Etude exp\'erimentale de la turbulence
  d'ondes \`a la surface d'un fluide. La th\'eorie de la Turbulence Faible \`a
  l'\'epreuve de la r\'ealit\'e pour les ondes de capillarit\'e et
  gravit\'e}},\ \href@noop {} {Ph.D. thesis},\ \bibinfo  {school} {Universit\'e
  Grenoble Alpes} (\bibinfo {year} {2016})\BibitemShut {NoStop}%
\bibitem [{\citenamefont {Kralchevsky}\ and\ \citenamefont
  {Nagayama}(2000)}]{Kralchevsky2000}%
  \BibitemOpen
  \bibfield  {author} {\bibinfo {author} {\bibfnamefont {Peter~A.}\
  \bibnamefont {Kralchevsky}}\ and\ \bibinfo {author} {\bibfnamefont {Kuniaki}\
  \bibnamefont {Nagayama}},\ }\bibfield  {title} {\enquote {\bibinfo {title}
  {{Capillary interactions between particles bound to interfaces, liquid films
  and biomembranes}},}\ }\href {\doibase 10.1016/S0001-8686(99)00016-0}
  {\bibfield  {journal} {\bibinfo  {journal} {Adv. Colloid Interface Science}\
  }\textbf {\bibinfo {volume} {85}},\ \bibinfo {pages} {145--192} (\bibinfo
  {year} {2000})}\BibitemShut {NoStop}%
\bibitem [{\citenamefont {Gifford}\ and\ \citenamefont
  {Scriven}(1971)}]{Gifford1971}%
  \BibitemOpen
  \bibfield  {author} {\bibinfo {author} {\bibfnamefont {W.~A.}\ \bibnamefont
  {Gifford}}\ and\ \bibinfo {author} {\bibfnamefont {L.~E.}\ \bibnamefont
  {Scriven}},\ }\bibfield  {title} {\enquote {\bibinfo {title} {{On the
  attraction of floating particles}},}\ }\href {\doibase
  10.1016/0009-2509(71)83003-8} {\bibfield  {journal} {\bibinfo  {journal}
  {Chemical Engineering Science}\ }\textbf {\bibinfo {volume} {26}},\ \bibinfo
  {pages} {287--297} (\bibinfo {year} {1971})}\BibitemShut {NoStop}%
\bibitem [{\citenamefont {Tayfun}(1980)}]{Tayfun1980}%
  \BibitemOpen
  \bibfield  {author} {\bibinfo {author} {\bibfnamefont {M.~Aziz}\ \bibnamefont
  {Tayfun}},\ }\bibfield  {title} {\enquote {\bibinfo {title} {{Narrow-band
  nonlinear sea waves}},}\ }\href {\doibase 10.1029/JC085iC03p01548} {\bibfield
   {journal} {\bibinfo  {journal} {J. Geophys. Res. : Oceans}\ }\textbf
  {\bibinfo {volume} {85}},\ \bibinfo {pages} {1548--1552} (\bibinfo {year}
  {1980})}\BibitemShut {NoStop}%
\bibitem [{\citenamefont {Socquet-Juglard}\ \emph {et~al.}(2005)\citenamefont
  {Socquet-Juglard}, \citenamefont {Dysthe}, \citenamefont {Trulsen},
  \citenamefont {Krogstad},\ and\ \citenamefont {Liu}}]{Socquet-Juglard2005}%
  \BibitemOpen
  \bibfield  {author} {\bibinfo {author} {\bibfnamefont {Herv{\'{e}}}\
  \bibnamefont {Socquet-Juglard}}, \bibinfo {author} {\bibfnamefont
  {Kristian~B.}\ \bibnamefont {Dysthe}}, \bibinfo {author} {\bibfnamefont
  {Karsten}\ \bibnamefont {Trulsen}}, \bibinfo {author} {\bibfnamefont
  {Harald~E.}\ \bibnamefont {Krogstad}}, \ and\ \bibinfo {author}
  {\bibfnamefont {Jingdong}\ \bibnamefont {Liu}},\ }\bibfield  {title}
  {\enquote {\bibinfo {title} {{Probability distributions of surface gravity
  waves during spectral changes}},}\ }\href {\doibase
  10.1017/S0022112005006312} {\bibfield  {journal} {\bibinfo  {journal} {J.
  Fluid Mech.}\ }\textbf {\bibinfo {volume} {542}},\ \bibinfo {pages}
  {195--216} (\bibinfo {year} {2005})}\BibitemShut {NoStop}%
\bibitem [{\citenamefont {Onorato}\ \emph {et~al.}(2009)\citenamefont
  {Onorato}, \citenamefont {Cavaleri}, \citenamefont {Fouques}, \citenamefont
  {Gramstad}, \citenamefont {Janssen}, \citenamefont {Monbaliu}, \citenamefont
  {Osborne}, \citenamefont {Pakozdi}, \citenamefont {Serio}, \citenamefont
  {Stansberg}, \citenamefont {Toffoli},\ and\ \citenamefont
  {Trulsen}}]{Onorato2009}%
  \BibitemOpen
  \bibfield  {author} {\bibinfo {author} {\bibfnamefont {M.}~\bibnamefont
  {Onorato}}, \bibinfo {author} {\bibfnamefont {L.}~\bibnamefont {Cavaleri}},
  \bibinfo {author} {\bibfnamefont {S.}~\bibnamefont {Fouques}}, \bibinfo
  {author} {\bibfnamefont {O.}~\bibnamefont {Gramstad}}, \bibinfo {author}
  {\bibfnamefont {Peter~A.E.M.}\ \bibnamefont {Janssen}}, \bibinfo {author}
  {\bibfnamefont {J.}~\bibnamefont {Monbaliu}}, \bibinfo {author}
  {\bibfnamefont {A.~R.}\ \bibnamefont {Osborne}}, \bibinfo {author}
  {\bibfnamefont {C.}~\bibnamefont {Pakozdi}}, \bibinfo {author} {\bibfnamefont
  {M.}~\bibnamefont {Serio}}, \bibinfo {author} {\bibfnamefont {C.~T.}\
  \bibnamefont {Stansberg}}, \bibinfo {author} {\bibfnamefont {a.}~\bibnamefont
  {Toffoli}}, \ and\ \bibinfo {author} {\bibfnamefont {K.}~\bibnamefont
  {Trulsen}},\ }\bibfield  {title} {\enquote {\bibinfo {title} {{Statistical
  properties of mechanically generated surface gravity waves: a laboratory
  experiment in a three-dimensional wave basin}},}\ }\href {\doibase
  10.1017/S002211200900603X} {\bibfield  {journal} {\bibinfo  {journal} {J.
  Fluid Mech.}\ }\textbf {\bibinfo {volume} {627}},\ \bibinfo {pages} {235}
  (\bibinfo {year} {2009})}\BibitemShut {NoStop}%
\bibitem [{\citenamefont {Leckler}\ \emph {et~al.}(2015)\citenamefont
  {Leckler}, \citenamefont {Ardhuin}, \citenamefont {Peureux}, \citenamefont
  {Benetazzo}, \citenamefont {Bergamasco},\ and\ \citenamefont
  {Dulov}}]{Leckler2015}%
  \BibitemOpen
  \bibfield  {author} {\bibinfo {author} {\bibfnamefont {Fabien}\ \bibnamefont
  {Leckler}}, \bibinfo {author} {\bibfnamefont {Fabrice}\ \bibnamefont
  {Ardhuin}}, \bibinfo {author} {\bibfnamefont {Charles}\ \bibnamefont
  {Peureux}}, \bibinfo {author} {\bibfnamefont {Alvise}\ \bibnamefont
  {Benetazzo}}, \bibinfo {author} {\bibfnamefont {Filippo}\ \bibnamefont
  {Bergamasco}}, \ and\ \bibinfo {author} {\bibfnamefont {Vladimir}\
  \bibnamefont {Dulov}},\ }\bibfield  {title} {\enquote {\bibinfo {title}
  {{Analysis and Interpretation of Frequency-Wavenumber Spectra of Young Wind
  Waves}},}\ }\href {\doibase 10.1175/JPO-D-14-0237.1} {\bibfield  {journal}
  {\bibinfo  {journal} {J. Phys. Ocean.}\ }\textbf {\bibinfo {volume} {45}},\
  \bibinfo {pages} {2484----2496} (\bibinfo {year} {2015})}\BibitemShut
  {NoStop}%
\bibitem [{\citenamefont {Taklo}\ \emph {et~al.}(2015)\citenamefont {Taklo},
  \citenamefont {Trulsen}, \citenamefont {Gramstad}, \citenamefont {Krogstad},\
  and\ \citenamefont {Jensen}}]{Taklo1}%
  \BibitemOpen
  \bibfield  {author} {\bibinfo {author} {\bibfnamefont {Tore Magnus~A}\
  \bibnamefont {Taklo}}, \bibinfo {author} {\bibfnamefont {Karsten}\
  \bibnamefont {Trulsen}}, \bibinfo {author} {\bibfnamefont {Odin}\
  \bibnamefont {Gramstad}}, \bibinfo {author} {\bibfnamefont {Harald~E}\
  \bibnamefont {Krogstad}}, \ and\ \bibinfo {author} {\bibfnamefont {Atle}\
  \bibnamefont {Jensen}},\ }\bibfield  {title} {\enquote {\bibinfo {title}
  {{Measurement of the dispersion relation for random surface gravity
  waves}},}\ }\href@noop {} {\bibfield  {journal} {\bibinfo  {journal} {Journal
  Of Fluid Mechanics}\ }\textbf {\bibinfo {volume} {766}},\ \bibinfo {pages}
  {326--336} (\bibinfo {year} {2015})}\BibitemShut {NoStop}%
\bibitem [{\citenamefont {Taklo}\ \emph {et~al.}(2017)\citenamefont {Taklo},
  \citenamefont {Trulsen}, \citenamefont {Krogstad},\ and\ \citenamefont
  {Nieto~Borge}}]{Taklo2}%
  \BibitemOpen
  \bibfield  {author} {\bibinfo {author} {\bibfnamefont {Tore Magnus~A}\
  \bibnamefont {Taklo}}, \bibinfo {author} {\bibfnamefont {Karsten}\
  \bibnamefont {Trulsen}}, \bibinfo {author} {\bibfnamefont {Harald~E}\
  \bibnamefont {Krogstad}}, \ and\ \bibinfo {author} {\bibfnamefont
  {Jos{\'e}~Carlos}\ \bibnamefont {Nieto~Borge}},\ }\bibfield  {title}
  {\enquote {\bibinfo {title} {{On dispersion of directional surface gravity
  waves}},}\ }\href@noop {} {\bibfield  {journal} {\bibinfo  {journal} {Journal
  Of Fluid Mechanics}\ }\textbf {\bibinfo {volume} {812}},\ \bibinfo {pages}
  {681--697} (\bibinfo {year} {2017})}\BibitemShut {NoStop}%
\bibitem [{\citenamefont {Collis}\ \emph {et~al.}(1998)\citenamefont {Collis},
  \citenamefont {White},\ and\ \citenamefont {Hammond}}]{Collis1998}%
  \BibitemOpen
  \bibfield  {author} {\bibinfo {author} {\bibfnamefont {W.B.}\ \bibnamefont
  {Collis}}, \bibinfo {author} {\bibfnamefont {P.R.}\ \bibnamefont {White}}, \
  and\ \bibinfo {author} {\bibfnamefont {J.K.}\ \bibnamefont {Hammond}},\
  }\bibfield  {title} {\enquote {\bibinfo {title} {{Higher-order spectra: the
  bispectrum and trispectrum}},}\ }\href {\doibase 10.1006/mssp.1997.0145}
  {\bibfield  {journal} {\bibinfo  {journal} {Mechanical Systems and Signal
  Processing}\ }\textbf {\bibinfo {volume} {12}},\ \bibinfo {pages} {375--394}
  (\bibinfo {year} {1998})}\BibitemShut {NoStop}%
\bibitem [{\citenamefont {V}(1994)}]{Krasit}%
  \BibitemOpen
  \bibfield  {author} {\bibinfo {author} {\bibfnamefont {Krasitskii}\
  \bibnamefont {V}},\ }\bibfield  {title} {\enquote {\bibinfo {title} {{On
  reduced equations in the Hamiltonian theory of weakly nonlinear surface
  waves}},}\ }\href@noop {} {\bibfield  {journal} {\bibinfo  {journal} {Journal
  Of Fluid Mechanics}\ }\textbf {\bibinfo {volume} {272}},\ \bibinfo {pages}
  {1--20} (\bibinfo {year} {1994})}\BibitemShut {NoStop}%
\end{thebibliography}%

\end{document}